\documentclass[a4paper,12pt]{article}
\usepackage{citeref}

\input cohmacros.sty

\advance\textheight3cm        
\advance\topmargin-2.0cm
\advance\textwidth2.6cm
\advance\evensidemargin-1.6cm
\advance\oddsidemargin-1.6cm

\def\Sb{{\bf S}} 

\begin{document}

\vspace*{-1.5cm}
\begin{center}
{\LARGE \bf Foundations of quantum physics} \\[4mm]

{\LARGE \bf III. Measurement} \\

\vspace{0.5cm}
\vspace{0.5cm}

\centerline{\sl {\large \bf Arnold Neumaier}}

\vspace{0.5cm}

\centerline{\sl Fakult\"at f\"ur Mathematik, Universit\"at Wien}
\centerline{\sl Oskar-Morgenstern-Platz 1, A-1090 Wien, Austria}
\centerline{\sl email: Arnold.Neumaier@univie.ac.at}
\centerline{\sl \url{http://www.mat.univie.ac.at/~neum}}

\end{center}



\hfill April 24, 2019
\vspace{0.5cm}

\bigskip
\bfi{Abstract.}
This paper presents the measurement problem from the point of view of 
the thermal interpretation of quantum physics introduced in Part II. 
Unlike most work on the foundations of quantum mechanics, the present 
paper involves a multitude of connections to the actual practice of 
quantum theory and quantum measurement.

The measurement of a Hermitian quantity $A$ is regarded as giving an 
uncertain value approximating the q-expectation $\<A\>$ rather than 
(as tradition wanted to have it) as an exact revelation of an 
eigenvalue of $A$. Single observations of microscopic systems are 
(except under special circumstances) very uncertain measurements only.

The thermal interpretation 
\\
\pt
treats detection events as a statistical measurement of particle beam 
intensity;
\\
\pt
claims that the particle concept is only asymptotically valid, under 
conditions where particles are essentially free.
\\
\pt 
claims that the unmodeled environment influences the results enough to 
cause all randomness in quantum physics.
\\
\pt
allows one to derive Born's rule for scattering and in the limit of 
ideal measurements; but in general, only part of Born's rule holds 
exactly: Whenever a quantity $A$ with zero uncertainty is measured 
exactly, its value is an eigenvalue of $A$;
\\
\pt
has no explicit collapse -- the latter emerges approximately in
non-isolated subsystems;
\\
\pt 
gives a valid interpretation of systems modeled by a quantum-classical 
dynamics;
\\
\pt
explains the peculiar features of the Copenhagen interpretation 
(lacking realism between measurements) and the minimal statistical 
interpretation (lacking realism for the single case) in the microscopic
domain where these interpretations apply.

The thermal interpretation is an interpretation of quantum physics that 
is in principle refutable by theoretical arguments leading to a negative
answer to a number of open issues collected at the end of the paper,
since there is plenty of experimental evidence for each of the points 
mentioned there.

\vfill

For the discussion of questions related to this paper, please use
the discussion forum \\
\url{https://www.physicsoverflow.org}.

\newpage
\tableofcontents 

\vspace{2cm}
\section{Introduction}

This paper presents the measurement problem from the point of view of 
the \bfi{thermal interpretation} of quantum physics introduced in 
Part II \cite{Neu.IIfound} of this series.

In the thermal interpretation of quantum physics, the theoretical 
observables (the beables in the sense of \sca{Bell} \cite{Bell})
are the expectations and functions of them. They satisfy a deterministic
dynamics. Some of these beables are practically (approximately) 
observable. In particular, q-expectations\footnote{\label{f.q}
As in Part I \cite{Neu.Ifound}  I follow the convention of 
\sca{Allahverdyan} et al. \cite{AllBN2}, and add the prefix 'q-' to all 
traditional quantum notions that have in the thermal view a new 
interpretation and hence a new terminology. 
In particular, we use the terms q-observable, q-expectation, q-variance,
q-standard deviation, q-probability, q-ensemble for the conventional 
terms observable, expectation, variance, standard deviation, 
probability, and ensemble.
} 
of Hermitian quantities and
q-probabilities, the probabilities associated with appropriate 
self-adjoint Hermitian quantities, are among the theoretical 
observables. The q-expectations are approximately measured by
reproducible single measurements of macroscopic quantities, or by
sample means in a large number of observations on independent, similarly
prepared microscopic systems. The q-probabilities are approximately
measured by determining the relative frequencies of corresponding 
events associated with a large number of independent, similarly 
prepared systems.

This eliminates all foundational problems that were introduced into 
quantum physics by basing the foundations on an unrealistic concept of
observables and measurement. With the thermal interpretation, the
measurement problem turns from a philosophical riddle into a
scientific problem in the domain of quantum statistical mechanics,
namely how the quantum dynamics correlates macroscopic readings from
an instrument with properties of the state of a measured microscopic
system. 

This is the subject of the present paper. Everything physicists
measure is measured in a thermal environment for which statistical
thermodynamics is relevant. The thermal interpretation agrees with how 
one interprets measurements in thermodynamics, the macroscopic part of 
quantum physics, derived via statistical mechanics.
By its very construction, the thermal interpretation
naturally matches the classical properties of our quantum world:
The thermal interpretation assigns states -- and a realistic 
interpretation for them -- to individual quantum systems, in a way that 
large quantum systems are naturally described by classical observables. 

Section \ref{s.meas} postulates a measurement principle that defines 
what it means in the thermal interpretation to measure a quantity with 
a specified uncertainty and discusses the role played by macroscopic 
systems and the weak law of large numbers in getting readings with 
small uncertainty. Since quantum physics makes makes many deterministic 
predictions, for example regarding observed spectra, but also assertions
about probabilities, we distinguish deterministic and statistical 
measurements.

Unlike in traditional interpretations, single, nonreproducible 
observations do not count as measurements since this would violate the 
reproducibility of measurements -- the essence of scientific practice.
As a consequence, the measurement of a Hermitian quantity $A$ is 
regarded as giving an uncertain value approximating the q-expectation 
$\<A\>$ rather than (as tradition wanted to have it) as an exact 
revelation of an eigenvalue of $A$. This difference is most conspicuous 
in the interpretation of single discrete microscopic events. Except in 
very special circumstances, these are not reproducible. Thus they have 
no scientific value in themselves and do not constitute measurement 
results. Scientific value is, however, in ensembles of such 
observations, which result in approximate measurements of 
q-probabilities and q-expectations.

Since relativistic quantum field theory is the fundamental theory of 
elementary particles and fields, with the most detailed description, 
the simpler quantum mechanics of particles is necessarily a derived 
description. Section \ref{s.partFields} discusses the extent to which a 
particle picture of matter and radiation is appropriate -- namely in 
scattering processes, where particles may be cosidered to be essentially
free except during a very short interaction time, and in the domain 
where the geometric optics perspective applies.

In 1852, at a time when Planck, Einstein, Bohr, Heisenberg, 
Schr\"odinger, Born, Dirac, and von Neumann -- the founders of modern 
quantum mechanics --  were not even born, George Stokes described all 
the modern quantum phenomena of a single qubit, explaining them in 
classical terms. Remarkably, this description of a qubit (recounted 
in Subsection \ref{ss.qubit}) is fully consistent with the
thermal interpretation of quantum physics. Stokes' description is
coached in the language of optics -- polarized light was the only
quantum system that, at that time, was both accessible to experiment
and quantitatively understood. Stokes' classical observables are
the functions of the components of the coherence matrix, the optical
analogue of the density operator of a qubit, just as the thermal
interpretation asserts.

Section \ref{s.tiStatMech} gives the thermal interpretation of 
statistical mechanics. All statistical mechanics is based
on the concept of coarse-graining, introduced in Subsection 
\ref{ss.coarse}. Due to the neglect of high frequency details, 
coarse-graining leads to stochastic features, either in the models 
themselves, or in the relation between models and reality. 
Deterministic coarse-grained models are usually chaotic, introducing a 
second source of randomness. The most important form of coarse-graining 
leads to statistical thermodynamics of equilibrium and nonequilibrium, 
leading for example to the Navier--Stokes equations of fluid mechanics. 
Other ways of coarse-graining lead to quantum-classical models, 
generalizing the Born--Oppenheimer approximation widely used in quantum 
chemistry.

A multitude of interpretations of quantum mechanics exist; most of them 
in several variants. As we shall see in Section \ref{s.trad}, the 
mainstream interpretations may be regarded as partial versions of the 
thermal interpretation. 
In particular, certain puzzling features of both the Copenhagen 
interpretation and the statistical interpretation get their explanation 
through the thermal interpretation of quantum field theory.
We shall see that these peculiar features get their natural 
justification in the realm for which they were created -- the 
statistics of few particle scattering events.

\bigskip

The bulk of this paper is intended to be nontechnical and understandable
for a wide audience being familiar with some traditional quantum
mechanics. The knowledge of some basic terms from functional analysis is
assumed; these are precisely defined in many mathematics books.
However, a number of remarks are addressed to experts and then refer to
technical aspects explained in the references given.

In the bibliography, the number(s) after each reference give the page 
number(s) where it is cited.

\bigskip
{\bf Acknowledgments.}
Earlier versions of this paper benefitted from discussions with
Rahel Kn\"opfel.

\section{The thermal interpretation of measurement}
\label{s.meas}

To clarify the meaning of the concept of measurement we postulate in 
Subsection \ref{ss.measurement} a measurement principle that defines 
what it means in the thermal interpretation to measure a quantity with 
a specified uncertainty.

The essence of scientific practice is the reproducibility of 
measureemnts, discussed in Subsection \ref{ss.SDmeas}. The next two 
subsections distinguish deterministic and statistical measurements 
depending on whether a single observation is reproducible, and discuss
the role played by macroscopic systems and the weak law of large 
numbers in getting readings with small uncertainty. The special case 
of event-based measurements described in terms of POVMs is considered 
in Subsection \ref{ss.event}.

\subsection{What is a measurement?}\label{ss.measurement}

According to the thermal interpretation, properties of the system to be 
measured are encoded in the state of the system and its dynamics.
This state and what can be deduced from it are the only objective
properties of the system. On the other hand, a measuring instrument 
measures properties of a system of interest. The measured value -- a 
pointer reading, a sound, a counter value, etc. -- is read off from the 
instrument, and hence is primarily a property of the measuring 
instrument and not one of the measured system. From the properties of 
the instrument (the instrument state), one can measure or compute the 
measurement results. Measurements are possible only if the microscopic 
laws imply quantitative relations between properties of the measured 
system (i.e., the system state) and the values read off from the 
measuring instrument (properties of the detector state).

This -- typically somewhat uncertain -- relation was specified 
in the rule (M) from Subsection 4.2 of Part I \cite{Neu.Ifound} 
that we found necessary for a good interpretation:

{\bf (M)} 
We say that a property $P$ of a system $S$ (encoded in its state) has 
been \bfi{measured} by another system, the \bfi{detector} $D$, if at the
time of completion
of the measurement and a short time thereafter (long enough that the
information can be read by an observer) the detector state carries
enough information about the state of the measured system $S$ at the
time when the measurement process begins to deduce with sufficient
reliability the validity of property $P$ at that time.

To give a precise formal expression for rule (M) in the context of the 
thermal interpretation, we have to define the property
$P$ as the validity or invalidity of a specific mathematical statement
$P(\rho_S)$ about the state $\rho_S$ of the system and the information
to be read as another specific mathematical statement $Q(\rho_D)$ about
the state $\rho_D$ of the detector. Then we have to check (theoretically
or experimentally) that the dynamics of the joint system composed of
system, detector, and the relevant part of the environment implies that,
with high confidence and an appropriate accuracy,
\lbeq{e.rhoDS}
Q(\rho_D(t)) \approx P(\rho_S(t_i)) \for t_f\le t\le t_f+\Delta t.
\eeq
Here $t_i$ and $t_f$ denote the initial and final time of the duration
of the measurement process, and $\Delta t$ is the time needed to read
the result.

For example, to have sufficient reasons to call the observation of a 
pointer position or a detector click an observation of a physical
property of the measured system one must show that \gzit{e.rhoDS}
holds for some encoding of the pointer position or detector click as
$Q(\rho_B)$ and the property $P(\rho_S)$ claimed to be measured.

Establishing such a relation \gzit{e.rhoDS} based on experimental 
evidence requires knowing already how system properties are 
experimentally defined, through preparation or measurement. This gives
the definition of measurement the appearance of a self-referential 
cycle, unless we can give an independent definition of preparation.
We shall come back to this later in Section \ref{s.trad}.

On the other hand, deducing \gzit{e.rhoDS} theoretically is a difficult 
task of statistical mechanics, since the instrument is a macroscopic 
body that, on the fundamental level necessary for a foundation, can be 
treated only in terms of statistical mechanics. The investigation of
this in Subsections \ref{ss.Balian} and \ref{ss.discrete} will show 
essential differences between the traditional interpretations and the 
thermal interpretation. 

Taking $P(\rho)=\tr\rho A=\<A\>$, weget as special case the following 
principle. It defines, in agreement with the general uncertainty 
principle (GUP) and todays NIST standard for specifying uncertainty 
(\sca{Taylor \& Kuyatt} \cite{TayK}) what it means to have measured a 
quantity:

\bfi{(MP)} \bfi{Measurement principle:}
{\it A macroscopic quantum device qualifies as an instrument for 
approximately, with uncertainty $\Delta a$, measuring a Hermitian 
quantity $A$ of a system with density operator $\rho$, if it satisfies
the following two conditions:
\\
(i) (\bfi{uncertainty}) All measured results $a$ deviate from $\ol A$ 
by approximately $\Delta a$. The measurement uncertainty is bounded 
below by $\Delta a\ge \sigma_A$.
\\
(ii) (\bfi{reproducability}) If the measurement can be sufficiently 
often repeated on systems with the same or a sufficiently similar state 
then the sample mean of $(a-\ol A)^2$ approaches $\Delta a^2$. 
} 

As customary, one writes the result of a measurement as an
\bfi{uncertain number} $a \pm\Delta a$ consisting of the measured value
value $a$ and its uncertainty deviation $\Delta a$, with the meaning 
that the error $|a-\ol A|$ is at most a small multiple of $\Delta a$.
Because of possible systematic errors, it is generally not possible
to interpret $a$ as mean value and $\Delta a$ as standard
deviation. Such an interpretation is valid only if the instrument
is calibrated to be unbiased.

The measurement principle (MP) creates the foundation of measurement 
theory. Physicists doing quantum physics (even those adhering to the
shut-up-and-calculate mode of working) use this rule routinely and
usually without further justification. The rule applies universally.
No probabilistic interpretation is needed.
In particular, the first part applies also to single measurements of
single systems.

The validity of the measurement principle for a given instrument must 
either be derivable from quantum models of the instrument by a 
theoretical analysis, or it must be checkable by experimental evidence
by calibration. 
In general, the theoretical analysis leads to difficult problems in
statistical mechanics that can be solved only approximately, and only in
idealized situations. From such idealizations one then transfers
insight to make educated guesses in cases where an analysis is too
difficult, and adjusts parameters in the design of the instrument
by an empirical calibration process.

\bigskip 

Consistent with the general uncertainty principle (GUP), the 
measurement principle (MP) demands that any instrument for measuring 
a quantity $A$ has an uncertainty $\Delta a\ge \sigma_A$.\footnote{ 
The formulation ''at least of the order of $\sigma_A$'' allows for the
frequent situation that the measurement uncertainty is larger than
the intrinsic (theoretical) uncertainty $\sigma_A$.
} 
It is an open problem how to prove this from the statistical mechanics 
of measurement models. But that such a limit cannot be overcome has 
been checked in the early days of quantum mechanics by a number of 
thought experiments. Today it is still consistent with experimental 
capabilities and no serious proposals exist that could possibly change 
this sitiation.

In particular, exact measurements have $\Delta a=0$ and hence 
$\sigma_A=0$.
This indeed happens for measurements of systems in a pure state when 
the state vector is an eigenstate of the quantity measured. Thus
part of Born's rule holds: {\it Whenever a quantity $A$ is measured 
exactly,\footnote{
But the discrete measurements in a Stern--Gerlach experiemnt, say, 
are not exact measurements in this sense, but very low accucacy
measurements of the associated q-expectations; see Subsection 
\ref{ss.event}.
} 
its value is an eigenvalue of $A$.} But for inexact (i.e, 
almost all) measurements, the thermal interpretation rejects Born's 
rule as an axiom defining what counts as a measurement result. 
With this move, all criticism from Part I \cite[Section 3]{Neu.Ifound} 
becomes void since Born's rule remains valid only in a limited validity;
see Subsection \ref{ss.Born}.

\subsection{Statistical and deterministic measurements}\label{ss.SDmeas}

The requirement (MP) for a measuring instrument includes the 
reproducibility of the resulting measurement values. Reproducibility in 
the general sense that all systems prepared in the same state have to 
behave alike when measured is a basic requirement for all natural 
sciences. The term ''alike'' has two different interpretations 
depending on the context: Either ''alike'' is
meant in the deterministic sense of ''approximately equal within the
specified accuracy''. Or ''alike'' is meant in the statistical sense of
''approximately reproducing in the long run the same probabilities and
mean values''. An object deserves the name ''instrument'' only if it
behaves in one or other of these ways. 

Corresponding to the two
meanings we distinguish two kinds of measuring instruments,
deterministic ones and statistical ones. Consequently, the quantitative 
relationship between the system state and the measurement results may 
be deterministic or statistical, depending on what is measured.

Radioactive decay, when modeled on the level of individual particles, 
is a typical \bfi{statistical} phenomenon. It needs a stochastic 
description as a branching process, similar to classical birth and 
death processes in biological population dynamics. The same holds for 
particle scattering, the measurement of cross sections, since particles 
may be created or annihilated, and for detection events, such as 
recording photons by a photoelectric device or particle trachs in a 
bubble chamber. 

On the other hand, although quantum physics generally counts as an 
intrinsically probabilistic theory, it is important to realize that it 
not only makes assertions about probabilities but also makes many
\bfi{deterministic} predictions verifiable by experiment.
These deterministic predictions fall into two classes:

(i) Predictions of numerical values believed to have a precise value
in nature:

\pt 
The most impressive proof of the correctness of quantum field theory in 
microphysics is the magnetic moment of the electron, predicted by
quantum electrodynamics (QED) to the phenomenal accuracy of 12
significant digit agreement with the experimental value.
It is a universal constant, determined solely by the two parameters
in QED, the electron mass and the fine structure constant.

\pt
QED also predicts correctly emission and absorption spectra
of atoms and molecules, both the spectral positions
and the corresponding line widths.

\pt
Quantum hadrodynamics allows the prediction of the masses of all
isotopes of the chemical elements in terms of models with only a
limited number of parameters.

(ii)  Predictions of qualitative properties, or of numerical values
believed to be not exactly determined but which are accurate with a
tiny, computable uncertainty.

\pt 
The modern form of quantum mechanics was discovered through its 
successful description and organization of a multitude of spectroscopic 
details on the position and width of spectral lines in atomic and 
molecular spectra.

\pt
QED predicts correctly the color of gold, the liquidity of
mercury at room temperature, and the hardness of diamond.

\pt
Quantum physics enables the computation of thermodynamic equations
of state for a huge number of materials. Equations of states are used 
in engineering in a deterministic manner, with negligible uncertainty. 
Engineers usually need not explicitly consider quantum effects since 
these are encoded in their empirical formulas for the equations of 
states.

\pt
quantum chemistry predicts correctly rates of chemical reactions.

\pt
From quantum physics one may also compute transport
coefficients for deterministic kinetic equations used in a variety
of applications.

Thus quantum physics makes both deterministic and statistical
assertions, depending on which system it is applied to and on the state
or the variables to be determined.
Statistical mechanics is mainly concerned with deterministic
prediction of class (ii) in the above classification.

Predictions of class (i) are partly related to spectral properties
of the Hamiltonian of a quantum system, and partly to properties deduced
from form factors, which are deterministic byproducts of scattering
calculations.
In both cases, classical measurements account adequately for the
experimental record.

The traditional interpretations of quantum mechanics do only 
rudimentarily address the deterministic aspects of quantum mechanics, 
requiring very idealized assumptions (being in an eigenstate of the 
quantity measured) that are questionable in all deterministic 
situations described above.

\subsection{Macroscopic systems and deterministic instruments}
\label{ss.macro}

A \bfi{macroscopic system} is a system large enough to be described 
sufficiently well by the methods of statistical mechanics,\footnote{
However, as discussed by \sca{Sklar} \cite{Skl}, both the frequentist 
and the subjective interpretation of probability in statistical 
mechanics have significant foundational problems, already in the 
framework of classical physics. These problems are absent in the thermal
interpretation, where single systems are described by mixed states, 
without any implied statistical connotation.
} 
where, due to the law of large numbers, one obtains essentially 
deterministic results. 

The weak law of large numbers implies that quantities averaged
over a large population of identically prepared systems become highly
significant when their value is nonzero, even when no single quantity 
is significant. This explains the success of Boltzmann's statistical 
mechanics to provide an effectively deterministic description of ideal 
gases, where all particles may be assumed to be independent and 
identically prepared.

In real, nonideal gases, the independence assumption is only
approximately valid because of possible interactions, and in liquids,
the independence is completely lost. The power of the statistical 
mechanics of Gibbs lies in the fact that it allows to replace simple 
statistical reasoning on populations based on independence by more 
sophisticated algebraic techniques that give answers even in extremely 
complex interacting cases.
Typically, the uncertainty is of the order $O(N^{-1/2})$, where $N$ is 
the mean number of identical microsystems making up the macroscopic 
system. Thus the thermal interpretation associates to macroscopic 
objects essentially classical quantities whose uncertain vaue 
(q-expectation) has a tiny uncertainty only. 

In particular, the macroscopic pointer of a measurement instrument
always has a well-defined position, given by the q-expectation of the 
Heisenberg operator $x(t)$ corresponding to the center of mass of its 
$N\gg 1$ particles at time $t$. The uncertain pointer position at time 
$t$ is $\< x(t)\>\pm\sigma_{x(t)}$, where the q-expectation is taken in 
the Heisenberg state of the universe (or any sufficiently isolated 
piece of it). Thus the position is fully determined by the state of the 
pointer -- but it is an uncertain position. By the law of large numbers,
the uncertainty $\sigma_{x(t)}$ is of order $N^{-1/2}$. Typically, this 
limit accuracy is much better than the accuracy of the actual reading. 
Thus we get well-defined pointer readings, leading within the reading 
accuracy to deterministic measurement results. 

Whether by this or by other means, whennever one obtains an essentially 
deterministic measurement result, we may say that measuring is done by 
a deterministic instrument:

A \bfi{deterministic instrument} is a measuring instrument that
measures beables, deterministic functions $F(\rho)$ of the state $\rho$
of the system measured, within some known margin of accuracy,
in terms of some property read from the instrument, a macroscopic 
system. 
A special case is the measurement of a quantity $A$, since the 
uncertain value $\ol A=\Tr \rho A$ of $A$ is a
function of the state $\rho$ of the system. Thus if measurements yield 
values $a\approx\ol A$ within some uncertainty $\Delta a$, the 
corresponding instrument is a deterministic instrument for
\bfi{measuring} $A$ within this accuracy.

\subsection{Statistical instruments}

The measurement of a tiny, \bfi{microscopic} system, often consisting
of only a single particle, is of a completely different nature.
Now the uncertainties do not benefit from the law of large numbers,
and the relevant quantities often are no longer significant, in the
sense that their uncertain value is already of the order of their
uncertainties.
In this case, the necessary quantitative relations between properties
of the measured system and the values read off from the measuring
instrument are only visible as stochastic correlations.

The results of single measurements are no longer reproducably
observable numbers. In the thermal interpretation, a single detection 
event is therefore not regarded as a measurement of a property of a
measured microscopic system, but only as a property of the macroscopic
detector correlated to the nature of the incident fields. 

This is the essential part where the thermal interpretation differs 
from tradition. Indeed, from a single detection event, one can only 
glean very little information about the state of a microscopic system.
Conversely, from the state of a microscopic system one can usually
predict only probabilities for single detection events.

All readings from a photographic image or from the scale of a measuring 
instrument, done by an observer, are deterministic measurements of an 
instrument property by the observer. Indeed, what is measured by the 
eye is the particle density of blackened silver on a photographic plate 
or that of iron of the tip of the pointer on the scale, and these are 
extensive variables in a continuum mechanical local equilibrium 
description of the instrument. 

The historically unquestioned interpretation of such detection events 
as the measurement of a particle position is one of the reasons for the 
failure of traditional interpretations to give a satisfying solution 
of the measurement problem. The thermal interpretation is here more 
careful and treats detection events instead as a statistical measurement
of particle beam intensity. 

To obtain comprehensive information about the state of a single
microscopic system is therefore impossible. To collect enough
information about the prepared state and hence the state of a
system measured, one needs either time-resolved measurements on a 
single stationary system (available, e.g., for atoms in ion traps or 
for electrons in quantum dots), or a population of identically prepared 
systems. In the latter case, one can get useful microscopic state 
infromation through quantum tomography, cf. Subsection \ref{ss.event}.

Thus in case of measurements on microscopic quantum systems, the 
quantitative relationship between measurement results and measured 
properties only takes the form of a statistical correlation.
The reproducably observable items, and hence the carrier of 
scientific information, are statistical mean values and probabilities. 
These are indeed predictable by quantum physics. But -- in contrast to 
the conventional terminology applied to single detection events for 
photons or electrons --  {\em the individual events no longer count as 
definite measurements of single system properties}. 

This characteristics of the thermal interpretation is an essential 
difference to traditional interpretations, for which each event is a 
definite measurement. 

A \bfi{statistical instrument} determines its final
measurement results from a large number of raw measurements by averaging
or by more advanced statistical procedures, often involving computer
processing. Again, due to the law of large numbers, one obtains
essentially deterministic results, but now from very noisy raw
measurements. Examples include low intensity photodetection, the
estimation of probabilities for classical or quantum stochastic
processes, astronomical instruments for measuring the properties of
galaxies, or the measurement of population dynamics in biology.

This behaviour guarantees reproducibility. In other words, systems
prepared in the same state behave in the same way under measurement --
in a deterministic sense for a deterministic instrument, and in a
statistical sense for a statistical one.
In both cases, the final measurement results approximate with a limited
accuracy the value of a function $F$ of the state of the system under
consideration.

\subsection{Event-based measurements}\label{ss.event}

Measurements in the form of discrete events (such as the appearance of
clicks, flashes, or particle tracks) may be described in terms of an
\bfi{event-based instrument} characterized by a discrete family of
possible measurement results  $a_1,a_2,\dots$ that may be real or
complex numbers, vectors, or fields, and nonnegative Hermitan
quantities  $P_1,P_2,\dots$ satisfying
\lbeq{e.Psum}
      P_1 + P_2 + \dots  = 1.
\eeq
The nonnegativity of the $P_k$ implies that all q-probabilities
\lbeq{e.Pprob}
      p_k = \<P_k\>=\Tr \rho P_k
\eeq
are nonnegative, and \gzit{e.Psum} guarantees that the q-probabilities
always add up to 1.
By its definition, the notion of q-probabilities belongs to the formal 
core of quantum mechanics and is independent of any interpretation. 

Unlike in all traditional interpretations, the thermal interpretation 
considers the observable result $a_k$ not as exact measurement results 
of some ''observable'' with counterintuitive quantum properties but as
a (due to the tiny sample size very low accucacy) statistical 
measurements of certain q-expectations.

In the thermal interpretation all q-expectations are beables; in
particular, all q-proba\-bilities are among the beables. As described 
in Part II \cite[Subsection 3.5]{Neu.IIfound}, a
q-probability $p$ may be approximately measured as relative frequency,
whenever there is an event-generating device (the \bfi{preparation}) 
that produces a large number $N$ of independent copies (realizations)
of the same quantum system. In this case, we requires that if the 
measured system is in the state $\rho$, the instrument gives the 
observable result $a_k$ with a relative frequency approaching the 
q-probability $p_k$ as the sample size gets arbitrarily large.

An \bfi{event-based instrument} is a statistical instrument measuring
the probability of events modeled by a discrete (classical or quantum)
statistical process. In the quantum case, it is mathematically
described by a \bfi{positive operator-valued measure}, short \bfi{POVM},
defined as a family $P_1,P_2,\dots$ of Hermitian, positive semidefinite
operators satsifying \gzit{e.Psum} (or a continuous generalization of
this).

POVMs originated around 1975 in work by \sca{Helstrom} \cite{Hel}
on quantum detection and estimation theory and are discussed in some
detail in \sca{Peres} \cite{Per}. They describe the most
general quantum measurement of interest in quantum information theory.
Which operators $P_k$ correctly describe a statistical instrument
can in principle be found out by suitable \bfi{calibration
measurements}. Indeed, if we feed the instrument with enough systems
prepared in known states $\rho_j$, we can measure approximate
probabilities $p_{jk}\approx\<P_k\>_j=\Tr \rho_j P_k$.
By choosing the states diverse
enough, one may approximately reconstruct $P_k$ from this information
by a process called \bfi{quantum tomography}. In quantum information
theory, the Hilbert spaces are finite-dimensional, hence the quantities
form the algebra $\Ez=\Cz^{N\times N}$ of complex $N\times N$ matrices.
In this case, the density operator is \bfi{density matrix} $\rho$, a
complex Hermitian $N\times N$-matrix with trace one, together with the
trace formula
\[
\<A\>= \Tr \rho A.
\]
Since $\<1\>=1$, a set of $N^2-1$ binary tests for specific states,
repeated often enough, suffices for the state determination. Indeed,
it is easy to see that repeated tests for the states $e^j$, the unit
vectors with just one entry one and other entries zero, tests the
diagonal elements of the density matrix, and since the trace is one,
one of these diagonal elements can be computed from the knowledge of
all others. Tests for $e^j+e^k$ and $e^j+ie^k$ for all $j<k$ then
allow the determination of the $(j,k)$ and $(k,j)$ entries.
Thus frequent  repetition of a total of $N-1+2{N\choose 2} = N^2-1$
particular tests determines the full state.
The optimal reconstruction to a given accuracy, using a minimal number
of individual measurements, is the subject of \bfi{quantum estimation
theory}, still an active frontier of research.

Distinguished from a stochastic instrument performing event-based 
measurements is an \bfi{event-based filter}, which turns an input state
$\rho$ with probability 
\[
p_k:=\<R_k^*R_k\>
\]
into an output state
\[
\rho_k:=\frac{1}{p_k} R_k\rho R_k^*.
\] 
Here the $R_k$ are operators satisfying 
\[
\sum_k R_k^*R_k=1.
\]
Which case occurred may be considered as an event; the collection of 
possible events is then described by the POVM with $P_k:= R_k^*R_k$.

\subsection{The thermal interpretation of eigenvalues}\label{ss.ev}

As discussed already in Part I \cite{Neu.Ifound},
the correspondence between observed values and eigenvalues is
only approximate, and the quality of the approximation improves with
reduced uncertainty. The correspondence is perfect only at zero
uncertainty, i.e., for completely sharp observed values.
To discuss this in detail, we need some results from functional
analysis. The \bfi{spectrum} $\spec A$\index{$\spec A$, spectrum} of a
linear operator on a Euclidean space $\Hz$ (a common domain of all 
relevant q-observables of a system) is the set of all
$\lambda\in\Cz$ for which no linear operator $R(\lambda)$ from the
completion $\ol \Hz$ of $\Hz$ to $\Hz$ exists such that
$(\lambda-A)R(\lambda)$ is the identity. $\spec A$ is always a closed
set.

A linear operator $A\in\Lin\Hz$ is called
\bfi{essentially self-adjoint} if it is
Hermitian and its spectrum is real (i.e., a subset of $\Rz$).
For $N$-level systems, where $\Hz$ is finite-dimensional, the spectrum
coincides with the set of eigenvalues, and every
Hermitian operator is essentially self-adjoint. In infinite dimensions,
the spectrum contains the eigenvalues, but not every number in the
spectrum must be an eigenvalue; and whether a Hermitian operator is
essentially self-adjoint is a question of correct boundary conditions.

{\bf Theorem.}
Let $A$ be essentially self-adjoint, with value $\ol A :=\<A\>$ and 
q-standard deviation $\sigma_A$ in a given state.
Then the spectrum of $A$ contains some real number $\lambda$ with
\lbeq{e.specabs}
|\lambda-\ol A|\le \sigma_A.
\eeq

\bepf
The linear operator $B=(A-\ol A)^2-\sigma_A^2$ is a quadratic function
of $A$, hence its spectrum consists of all
$\lambda':=(\lambda-\ol A)^2-\sigma_A^2$ with $\lambda\in \spec A$;
in particular, it is real. Put $\lambda_0:=\inf\,\spec B$. Then
$B-\lambda_0$ is a Hermitian operator with a real, nonnegative spectrum,
hence positive semidefinite. (In infinite dimensions, this requires
the use of the spectral theorem.) Thus $B-\lambda_0\ge 0$ and
$0\le \<B-\lambda_0\>=\<(A-\ol A)^2\>-\sigma_A^2-\lambda_0=-\lambda_0$.
Therefore $\lambda_0\le 0$. Since $\spec B$ is closed, $\lambda_0$ is
in the spectrum, hence has the form $(\lambda-\ol A)^2-\sigma_A^2$
with $\lambda\in \spec A$. This $\lambda$ satisfies \gzit{e.specabs}.
\epf

In particular, if, in some state, $A$ has a \bfi{sharp} observable 
value, defined by $\sigma_A=0$,
then the value $\<A\>$ belongs to the spectrum. In practice, this is
the case only for quantities $A$ whose spectrum (set of sharp values)
consists of rationals with small numerator and denominator.
Examples are spin and polarization in a given direction, (small)
angular momentum, and (small) particle numbers.

\section{Particles from quantum fields}\label{s.partFields}

In continuation of the discussion in Subsection 4.4 of Part I 
\cite{Neu.Ifound}, we discuss in this Section the extent to which a 
particle picture of matter and radiation is appropriate.

In physics practice, it is often unavoidable to switch between 
representations featuring different levels of detail. The fundamental 
theory of elementary particles and fields, with the most detailed 
description, is quantum field theory. Since quantum field theory is 
fundamental, the simpler quantum mechanics of particles is necessarily 
a derived description. 

How to obtain the quantum mechanics of particles from relativistic 
interacting quantum field theory is a nontrivial problem. The 
traditional textbook description in terms of scattering and associated 
propagators does not give a description at finite times.

In the fundamental reality -- i.e., represented by beables of quantum 
field theory, expressed at finite times in hydrodynamic terms --, fields
concentrated in fairly narrow regions move along uncertain flow lines 
determined by effective field equations. 

In the particle description, these fields are somehow replaced by 
a quantum mechanical model of moving particles. The uncertainty in now
accounted for by the uncertain value of the position $\q(t)$ of each 
particle together with its uncertainty $\sigma_{\q(t)}$, at any time 
$t$, providing not a continuous trajectory but a fuzzy world tube 
defining their location. The momentum of the quantum particles is also 
uncertain. For example, the momentum vector of a particle at CERN 
is measured by collecting information from many responding wires and 
applying curve fitting techniques to get an approximate curve of 
positions at all times and inferring from its derivative an uncertain 
momentum. Similar techniques are used for particle tracks on 
photographic plates or in bubble chambers.

How one finds from a relativistic quantum field description of a beam a 
corresponding quantum mechanical particle description has hardly 
received attention so far. While informally, particles are considered 
to be elementary excitations of the quantum fields, this can be given 
an exact meaning only for free field theories. In interacting 
relativistic quantum fields, the notion is, at finite times, 
approximate only.

That the approximation problem is nontrivial can be seen from 
the fact that in quantum field theory, position is a certain parameter. 
whereas in the quantum mechanics of particles, position is an uncertain 
quantity. Thus in the approximation process, position loses its 
parameter status and becomes uncertain. How, precisely, is unknown.

\subsection{Fock space and particle description}\label{ss.partFree}

A precise correspondence between particles and fields is possible only 
in free quantum field theories. These are described by 
distribution-valued operators on a Fock space. The latter is completely 
determined by its 1-particle sector, the single particle space.

Poincare invariance, locality, and the uniqueness of the 
vacuum state imply that the single particle space of a free quantum 
field theory furnishes a causal unitary irreducible representation of 
the Poincare group. These representations were classified 
in 1939 by \sca{Wigner} \cite{Wig}. This is why particle theorists say 
that elementary particles are causal unitary irreducible representations
of the Poincare group, Thus elementary particles are something 
exceedingly abstract, not tiny, fuzzy quantum balls! 

For spin $\le 1$, these representations happen to roughly match the
solution space of certain wave equations for a single relativistic
particle in the conventional sense of quantum mechanics, but only if
one discards the contributions of all negative energy states of the
latter. In relativistic quantum field theory, the latter reappear as
states for antiparticles -- a different kind of particles with 
different properties. This already shows that there is something very 
unnatural about the relativistic particle picture on the 
quantum-mechanical single-particle level. 

In general, a field description on the particle level in terms of a 
conventional multiparticle structure is necessarily based on a Fock 
space representation with a number operator $N$ with spectrum consisting
precisely of the nonnegative integers. The eigenspace for the 
eigenvalue $1$ of $N$ then defines the bare single-particle Hilbert 
space. In the relativistic case, the resulting description is one 
in terms of bare, unphysical particles. 

Untangling the S-matrix using bare perturbation theory replaces the 
real-time dynamics of the quantum fields by an non-temporal infinite 
sum of contributions of multivariate integrals depicted in shorthand 
by Feynman diagrams showing a web of virtual particles. The Feynman 
diagrams provide a pictorial representation of the formalism of bare
perturbation theory. Free real particles show as external lines, while 
the interaction is represented in terms of internal lines, figuratively 
called virtual particles. Most of the resulting integrals (all except 
the tree diagrams) are infinite and physically meaningless. 
A renormalization process turns the sum of all diagrams with a fixed 
number of loops (where the infinities cancel) into finite numbers whose 
sum over not too high orders (the series is asymptotic only) has an 
(approximate) physical meaning. But in the renormalization process
the intuitive connection of the lines depicted in Feynman diagrams -- 
the alleged world lines of virtual particles, in the popular myth 
(cf. \sca{Neumaier} \cite{Neu.myth}) -- gets completely lost. Nothing 
resembles anything like a process in time -- described by the theory 
and the computations is only a black box probabilistic model of the 
in-out behavior of multiparticle scattering.

\subsection{Physical particles in interacting field theories}

\nopagebreak
\hfill\parbox[t]{10.8cm}{\footnotesize

{\em All our knowledge concerning the internal properties of atoms is 
derived from experiments on their radiation or collision reactions,
such that the interpretation of experimental facts ultimately depends 
on the abstractions of radiation in free space, and free material 
particles. [...]\\
The use of observations concerning the behaviour of particles in the 
atom rests on the possibility of neglecting, during the process of 
observation, the interaction between the particles, thus regarding them 
as free. [...]\\ 
The wave mechanical solutions can be visualised only in so far as they 
can be described with the aid of the concept of free particles. [...]\\ 
Summarising, it might be said that the concepts of stationary states 
and individual transition processes within their proper field of 
application possess just as much or as little ' reality' as the
very idea of individual particles.}

\hfill Niels Bohr, 1927 \cite[pp.586--589]{Boh1927}
}

\bigskip

While the conventional construction of relativistic quantum field 
theories starts with Fock space, a relativistic interacting quantum 
field itself cannot be described cannot be described in terms of a 
Fock space. The Fock space structure of the initial scaffolding is 
destroyed by the necessary renormalization, since the
number operator cannot be renormalized. Only the asymptotic fields
figuring in the S-matrix reside in a Fock space -- for colored quarks
because of confinement not even in a conventional Fock space with a
positive definite inner product, but only in an indefinite Fock--Krein
space. 

As a consequence, the particle concept is only asymptotically 
valid, under conditions where particles are essentially free.
Traditionally, the discussion of particle issues in relativistic 
interacting quantum fields is therefore restricted to scattering 
processes involving asymptotical particle states. Only the S-matrix 
provides meaning to quantum particles, in an asymptotic sense, 
describing Born's rule for scattering processes. In the formulation of
Part I \cite[Subsection 3.1]{Neu.Ifound}:
In a scattering experiment described by the S-matrix $S$,
\[
\Pr(\psi_\out|\psi_\iin):=|\psi_\out^*S\psi_\iin|^2
\]
is the conditional probability density that scattering of particles
prepared in the in-state $\psi_\iin$ results in particles
in the out-state $\psi_\out$. 

Indeed, textbook scattering theory for elementary particles is the 
{\em only} place where Born's rule is used in quantum field theory. 
Here the in- and out-states are asymptotic eigenstates of total 
momentum, labelled by a maximal collection of independent quantum 
numbers (including particle momenta and spins). An \bfi{asymptotic 
quantity} is a q-observable still visible in the limits of time 
$t\to\infty$ or $t\to-\infty$, so that scattering theory says
something interesting about it. This is relevant since quantum dynamics
is very fast but measurements take time. Measuring times are already
very well approximated by infinity, on the time scale of typical quantum
processes. Thus only asymptotic quantities have a reasonably
well-defined response.
That's why information about microsystems is always collected via
scattering experiments described by the S-matrix, which connects
asymptotic preparation at time $t=-\infty$ with asymptotic measurement
at time $t=+\infty$.
Particle momenta (like other conserved additive quantities) are 
asymptotic quantities.

In quantum field theory, scattering theory is 
just the special case of a universe containing only a tiny number of 
particles with known momentum at time $t=-\infty$, whose behavior at 
time $t=+\infty$ is to be predicted. 
This caricature of a universe is justified only
when the few-particle system is reasonably well isolated from the
remainder of the universe. In a real experiment, this is a good
approximation to a collision experiment when the length and time scale
of a collision is tiny compared to the length and time scale of the
surrounding preparation and detection process. Much care is taken in
modern colliders to achieve this to the required degree of accuracy.

\subsection{Semiclassical approximation and geometric optics}
\label{ss.geomOpt}

In the preceding, we discussed the precise notion of particles in 
relativistic quantum field theory -- an asymptotic notion only. 
Cross sections for the scattering processes computed in this way are 
supposed to be exact (assuming ithe idealization that the underlying 
theory is exact and the computations are done exactly). 

However, the particle picture has another very practical use, as an 
approximate, semiclassical concept valid whenever the fields are 
concentrated along a single (possibly bent) ray and the resolution is 
coarse enough. 
When these conditions apply, one is no longer in the full quantum 
domain and can already describe everything semiclassically, i.e., 
classical with small quantum corrections. Thus the particle concept is 
useful when and only when the semiclassical description is already 
adequate. Whenever one uses the particle picture beyond scattering 
theory (and in particular always when one has to interpret what people 
using the particle language say), one silently acknowledges that one 
works in a semiclassical picture where a particle description makes 
approximate sense except during collisions. 

A \bfi{particle} is a blop of high field concentrations 
well-localized in phase space (i.e., in the kinetic approximation of 
quantum field theory), with a boundary whose width (or the width in 
transversal directions for a moving particle) is tiny compared 
to its diameter.

Thus field concentrations must be such
that their (smeared) density peaks at reasonably well-defined locations
in phase space. At this point, similar to the regime of geometric
optics for classical electromagnetic fields these peaks behave like
particles. Thus particles are approximately defined as local
excitations of a field, and they have (as wavelets in classical
mechanics) an uncertain (not exactly definable) position.
Their (necessarily approximate) position and momentum behaves
approximately classically (and gives rise to a classical picture of
quantum particles) in the regime corresponding to geometric
optics. When the spatial resolution is such that the conditions for the
applicability of geometric optics hold, particles can be used as an 
adequate approximate concept. 

In a collision experiment, it is valid to say that particles travel on 
incoming and outgoing beams in spacetime while they are far apart, 
since this is a good semiclassical description of the free particles 
in a paraxial approximation.
But when they come close, the semiclassical description breaks down 
and one needs full quantum field theory to describe what happens. 

The exact state of the interacting system is now a complicated state 
in a renormalized quantum field Hilbert space\footnote{
Because of superselection sectors, this Hilbert space is generally 
nonseparable, a direct sum of the Hilbert spaces corresponding to the 
different superselection sectors.
} 
that no one so far was able to characterize; it is only known
(Haag's theorem) that it cannot be the asymptotic Fock space describing 
the noninteracting particles. 
Since it is not a Fock space, talking about particles during the 
interaction makes no longer sense - the quantum fields of which the 
particles are elementary excitations become very non-particle like.
After the collision products separate well enough, the semiclassical 
description becomes feasible again, and one can talk again about 
particles traveling along beams. 

Thus while the field picture is always valid, the picture of particles
traveling along beams or other world tubes is appropriate except 
close to the collision of two world tubes. The behavior there is 
effectively described in a black box fashion by the S-matrix. This is a 
reasonable approximation if the collision speed is high enough, so 
that one can take the in- and outgoing particles as being at time 
$-\infty$ and $+\infty$, and can ignore what happens at finite times, 
i.e., during the encounter. Thus, in the \bfi{semiclassical} 
description, we have between collisions real particles described by 
asymptotic states, while the collisions themselves -- where the 
particle picture no longer make sense -- are described 
using a black box view featuring the S-matrix, To calculate the S-matrix
one may work in renormalized perturbation theory using quantum field 
theory.

Using the intuition of geometric optics requires a locally free
effective description.
In a locally homogeneous background, such an effective description is
usually achievable through the introduction of \bfi{quasiparticles}.
These are collective field modes that propagate as if they were free.
If the composition of the background changes, the definition of the
quasiparticles changes as well.

In particular, the photons in glass or air are quasiparticles 
conceptually different from those in vacuum. Similarly, the moving 
electrons in a metal are quasiparticles conceptually different from 
those in vacuum.
This shows that photons, electrons, and other elementary particles have
no conceptual identity across interfaces. A photon, traditionally
taken to be emitted by a source, then passing a system of lenses,
prisms, half-silvered mirrors, and other optical equipment, changes
its identity each time it changes its environment!

This is corroborated by the field of \bfi{electron optics},
where geometric rays are used to calculate properties of magnetic and
electrostatic lenses for electron beams.

\bigskip

Problems abound if one tries to push the analogies beyond the 
semiclassical domain of validity of the particle concept.
Already in classical relativistic mechanics, point trajectories are
idealizations, restricted to a treatment of the motion of a single
point in a classical external field. By a result of \sca{Currie} et al.
\cite{CurJS}, classical relativistic multi-particle point trajectories
are inconsistent with a Hamiltonian dynamics. Thus one should not
expect them to exist in quantum physics either. They are appropriate
only as an approximate description.

Note that this semiclassical domain of validity of the particle picture 
excludes experiments with multilocal fields generated by beam-splitters,
half-silvered mirrors, double slits, diffraction, long-distance 
entanglement, and the like. it is there where the attempt to stick to
the particle picture leads to all sorts of counterintuitive features.
But these are caused by the now inadequate particle imagery, not by 
strange features of quantum field theory itself.

\subsection{The photoelectric effect}

In quantum optics experiments, both sources and beams are extended
macroscopic objects describable by quantum field theory and statistical
mechanics, and hence have (according to the thermal interpretation)
associated nearly classical observables -- densities, intensities,
correlation functions -- computable from quantum physics in terms of
q-expectations.

An instructive example is the \bfi{photoelectric effect}, the
measurement of a classical free electromagnetic field by means of a
photomultiplier. A detailed discussion is given in Sections 9.1--9.5
of \sca{Mandel \& Wolf} \cite{ManW}; here we only give an informal
summary of their account.

Classical input to a quantum system is conventionally represented in
the Hamiltonian of the quantum system by an interaction term
containing the classical source as an external field or potential.
In the semiclassical analysis of the photoelectric effect, the
detector is modeled as a many-electron quantum system, while the
incident light triggering the detector is modeled as an external
electromagnetic field. The result of the analysis is that if the
classical field consists of electromagnetic waves (light) with a
frequency exceeding some threshold then the detector emits a random
stream of photoelectrons with a rate that, for not too strong light,
is proportional to the intensity of the incident light. The
predictions are quantitatively correct for normal light.

The response of the detector to the light is statistical,
and only the rate (a short time mean) with which the electrons are
emitted bears a quantitative relation with the intensity.
Thus the emitted photoelectrons form a statistical measurement
of the intensity of the incident light.

The results on this analysis are somewhat surprising: Although the
semiclassical model used to derive the quantitatively correct
predictions does not involve photons at all, the discrete nature of the
electron emissions implies that a photodetector responds
to classical light as if it were composed of randomly arriving photons!
(The latter was the basis for the original explanation of the 
photoeffect for which Einstein received the Nobel prize.)

This proves that the discrete response of a photodetector cannot be due
to the quantum nature of the detected object.

The classical external field discussed so far is of course only an 
approximation to the quantum electromagnetic field, and was only used 
to show that the discrete response of a photodetector cannot be due
to its interactions with particles, or more generally not to the 
quantum nature of the detected object. The discrete response is due to
the detector itself, and triggered by the interaction with a field. 
A field mediating the interaction must be present with sufficient 
intensity to transmit the energy necessary for the detection events. 
Both a classical and a quantum field produce such a response. Only the 
quantitiative details change in the case of quantum fields, but nothing 
depends on the presence or absence of ''photons''. Thus photons are 
figurative properties of quantum fields manifesting themselves only in 
the detectors. Before detection, there are no photons; one just has
 beams of light in an entangled state. 

This shows the importance of differentiating between prepared states
of the system (here of classical or quantum light) and measured events 
in the instrument (here the amplified emitted electrons). The 
measurement results are primarily a property of the instrument, and 
their interpretation as a property of the system measured needs 
theoretical analysis to be conclusive.

\subsection{A classical view of the qubit}\label{ss.qubit}

It is commonly said that quantum mechanics originated in 1900 with Max 
Planck, reached its modern form with Werner Heisenberg and Erwin 
Schr\"odinger, got its correct interpretation with Max Born, and its 
modern mathematical formulation with Paul Dirac and John von Neumann. 
It is very little known that much earlier -- in 1852, at a time when 
Planck, Heisenberg, Schr\"odinger, Born, Dirac, and von Neumann were 
not even born --, George Stokes described all the modern quantum 
phenomena of a single qubit, explaining them in classical terms.

Remarkably, this description of a qubit is fully consistent with the
thermal interpretation of quantum physics. Stokes' description is
coached in the language of optics -- polarized light was the only
quantum system that, at that time, was both accessible to experiment
and quantitatively understood. Stokes' classical observables are
the functions of the components of the coherence matrix, the optical
analogue of the density operator of a qubit, just as the thermal
interpretation asserts.

The transformation behavior of rays of completely polarized light
was first described in 1809 by Etienne-Louis \sca{Malus} \cite{Mal}
(who coined the name ''polarization''); that of partially polarized
light in 1852 by George {\sc Stokes} \cite{Sto}. This subsection gives
a modern description of the core of this work by Malus and Stokes.

We shall see that Stokes' description of a polarized quasimonochromatic
beam of classical light behaves exactly like a modern quantum bit.

A ray (quasimonochromatic beam) of polarized light of fixed frequency
is characterized by a state, described equivalently by a real
{\bf Stokes vector}
\[
S=(S_0,S_1,S_2,S_3)^T={S_0\choose \Sb}
\]
with
\[
S_0\ge |\Sb| = \sqrt{S_1^2+S_2^2+S_3^2},
\]
or by a {\bf coherence matrix}, a complex positive semidefinite
$2\times 2$ matrix $\rho$. These are related by
\[
\D\rho = \half(S_0 + \Sb \cdot\Bsigma)
=\half\pmatrix{S_0+S_3 & S_1-iS_2\cr S_1+iS_2 & S_0-S_3},
\]
where $\Bsigma$ is the vector of Pauli matrices.
$\Tr\rho=S_0$ is the {\bf intensity} of the beam.
$p=|\Sb|/S_0\in[0,1]$ is the \bfi{degree of polarization}.
Note the slight difference to density matrices, where the trace is
required to be one.

A linear, non-mixing (not depolarizing) instrument (for example a
polarizer or phase rotator) is characterized by a complex $2\times 2$
{\bf Jones matrix} $T$.
The instrument transforms an in-going beam in the state $\rho$
into an out-going beam in the state $\rho'=T\rho T^*$.
The intensity of a beam after passing the instrument is
$S_0'=\Tr \rho'=\Tr T\rho T^*=\Tr \rho T^*T$.
If the instrument is lossless, the intensities of the in-going
and the out-going beam are identical.
This is the case if and only if the Jones matrix $T$ is unitary.

Since $\det\rho=(S_0^2-S_3^2)-(S_1^2+S_2)^2=S_0^2-\Sb^2$,
the fully polarized case $p=1$, i.e., $S_0=|\Sb|$, is equivalent with
$\det\rho=0$, hence holds iff the rank of $\rho$ is $0$ or $1$. In this
case, the coherence matrix can be written in the form $\rho=\psi\psi^*$
with a state vector $\psi$ determined up to a phase. Thus precisely the
pure states are fully polarized. In this case, the intensity of the beam
is
\[
S_0=\<1\>=|\psi|^2=\psi^*\psi.
\]
A {\bf polarizer} has $T=\phi\phi^*$, where $|\phi|^2=1$.
It reduces the intensity to
\[
S_0'=\<T^*T\>=|\phi^*\psi|^2.
\]
This is {\bf Malus' law}.

An instrument with Jones matrix $T$ transforms a beam in the pure state
$\psi$ into a beam in the pure state $\psi'=T\psi$.
Passage through inhomogeneous media can be modeled by means of many
slices consisting of very thin instruments with Jones matrices $T(t)$
close to the identity.
If $\psi(t)$ denotes the pure state at time $t$ then
$\psi(t+\Delta t) = T(t) \psi(t)$, so that for small $\Delta t$
(the time needed to pass through one slice),
\[
\frac{d}{dt}\psi(t)
= \frac{\psi(t+\Delta t)-\psi(t)}{\Delta t}+O(\Delta t)
= \frac{(T(t)-1)}{\Delta t} \psi(t)+O(\Delta t).
\]
In a continuum limit $\Delta t\to 0$ we obtain the time-dependent
{\bf Schr\"odinger equation}
\[
i\hbar\frac{d}{dt}\psi(t) = H(t) \psi(t),
\]
where (note that $T(t)$ depends on $\Delta t$)
\[
H(t)=\lim_{\Delta t\to 0} i\hbar\frac{T(t)-1}{\Delta t}
\]
plays the role of a time-dependent Hamiltonian. Note that in the
lossless case, $T(t)$ is unitary, hence $H(t)$ is Hermitian.

A linear, mixing (depolarizing) instrument transforms $\rho$ instead
into a sum of several terms of the form $T\rho T^*$.
It is therefore described by a real $4\times 4$ {\bf Mueller matrix}
acting on the Stokes vector.
Equivalently, it is described by a completely positive linear map on
the space of $2\times 2$ matrices, acting on the polarization matrix.

Thus we see that a polarized quasimonochromatic beam of classical light
behaves exactly like a modern quantum bit. We might say that classical
optics is just the quantum physics of a single qubit passing through
a medium!

Indeed, the 1852 paper by \sca{Stokes} \cite{Sto} described all the
modern quantum phenomena for qubits, explained in classical terms.
In particular,

\pt
Splitting fully polarized beams into two such beams with different, but
orthogonal polarization corresponds to writing a wave function as
superposition of preferred basis vectors.

\pt
Mixed states are defined (in his paragraph 9) as arising from ''groups
of independent polarized streams'' and give rise to partially polarized
beams.

\pt
The coherence matrix is represented by Stokes with four real parameters,
in today's terms comprising the Stokes vector.

\pt
Stokes asserts (in his paragraph 16) the impossibility of recovering
from a mixture of several distinct pure states any information about
these states beyond what is encoded in the Stokes vector (equivalently,
the coherence matrix).

\pt
The latter can be linearly decomposed in many essentially distinct ways
into a sum of pure states, but all these decompositions are optically
indistinguishable, hence have no physical meaning.

The only difference to the modern description is that the microscopic 
view is missing.
For faint light, photodetection leads to discrete detection events
-- even in models with an external {\em classical} electromagnetic
field; cf. the discussion in Subsection \ref{s.partFields} below.
The trace of $\rho$ is the intensity of the beam, and the rate of
detection events is proportional to it. After normalization to unit
intensity, $\rho$ becomes the density operator corresponding to a
single detection event (aka photon).

This is a simple instance of the transition from a beam (classical
optics or quantum field) description to a single particle (quantum
mechanical) description.

It took 75 years after Stokes until the qubit made its next appearance 
in the literature, in a much less comprehensive way. In 1927,  
\sca{Weyl} \cite[pp.8-9]{Weyl1927} discusses qubits in the guise of an 
ensemble (''Schwarm'') of spinning electrons. Instead of the language 
of Stokes, the description uses the paradoxical language still in use 
today, where the meaning of everything must be redefined to give at 
least the appearance of making sense.

In its modern formulation via Maxwell's equations, classical partially 
polarized light (as described by Stokes) already requires the
stochastic form of these equations, featuring -- just like the full 
quantum description -- field expectations and correlation functions; 
see \sca{Mandel \& Wolf} \cite{ManW}. The coherence matrices turn into 
simple camatrix-valued field correlation functions.

\section{The thermal interpretation of statistical mechanics}
\label{s.tiStatMech}

Like quantum mechanics, quantum statistical mechanics also consists of
a formal core and its interpretation. Almost everything done in the
subject belongs to the formal core, the formal shut-up-and-calculate
part of statistical mechanics, without caring about the meaning of the
computed q-expectations. The interpretation is considered to be almost
obvious and hence gets very little attention. For example, the 
well-known statistical physics book by 
\sca{Landau \& Lifschitz} \cite{LL.5} dedicates just 7 (of over 500)
pages (in Section 5) to the properties of the density operator, the 
basic object in quantum statistical mechanics, and less than half of 
these pages concern its interpretation in terms of pure states. 
Fortunately, no use at all is made of this elsewhere in their book, 
since, as already discussed in Subsection 3.4 of Part I 
\cite{Neu.Ifound}, the ''derivation'' given there -- though one of the 
most carefully argued available in the literature -- is highly deficient
On the other hand, in their thermodynamic implications later in the 
book, they silently assume the thermal interpretation, by identifying 
(e.g., in Section 35, where they discuss the grand canonical ensemble) 
the thermodynamic energy and thermodynamic particle number with the 
q-expectation of the Hamiltonian and the number operator! 

The thermal interpretation revises the interpretation of quantum 
statistical mechanics and extends this revised interpretation
to the microscopic regime, thus accounting for the fact that there is no
clear boundary where the macroscopic becomes microscopic. Thus we do 
not need to assume anything special about the microscopic regime.

Subsection \ref{ss.Koopman} shows in which sense classical statistical 
mechanics is a special case of quantum statistical mechanics; thus it
suffices to discuss the quantum case. All statistical mechanics is based
on the concept of coarse-graining, introduced in Subsection 
\ref{ss.coarse}. Due to the neglect of high frequency details, 
coarse-graining leads to stochastic features, either in the models 
themselves, or in the relation between models and reality. 
Deterministic coarse-grained models are usually chaotic, introducing a 
second source of randomness, discussed in Subsection \ref{ss.Balian}.

Statistical mechanics proper starts with the discussion of Gibbs states 
(Subsection \ref{ss.Gibbs}) and the statistical thermodynamics of 
equilibrium and nonequilibrium (Subsection \ref{ss.noneq}). Other ways
of coarse-graining lead to quantum-classical models (Subsections 
\ref{ss.QCd} and \ref{ss.QCex}), generating among others the 
Born--Oppenheimer approximation widely used in quantum chemistry.

\subsection{Koopman's representation of classical statistical mechanics}
\label{ss.Koopman}

Classical mechanics can be written in a form that looks like quantum 
mechanics. Such a form was worked out by \sca{Koopman} \cite{Koo} for 
classical statistical mechanics. In the special case where one restricts
the expectation mapping to be a $*$-algebra homomorphism, all 
uncertainties vanish, and the Koopman representation describes 
deterministic classical Hamiltonian mechanics. 

We discuss classical statistical mechanics in terms of a commutative 
Euclidean $*$-algebra $\Ez$ of \bfi{ random variables}, i.e., Borel 
measurable complex-valued functions on a Hausdorff space $\Omega$, 
where bounded continuous functions are strongly integrable and the 
integral is given by $ \sint f := \int d\mu(X) f(X)$ for some 
distinguished measure $\mu$.
(For a rigorous treatment see \sca{Neumaier \& Westra} \cite{NeuW}.)
The quantities and the density operator $\rho$ are represented by 
multiplication operators in some Hilbert space of functions on phase 
space. The classical Hmailtonian $H(p,q)$ is replaced by the 
\bfi{Koopman Hamiltonian}
\[
\wh H:=\frac{\partial H(p,q)}{\partial q}i\frac{\partial}{\partial p}
       -\frac{\partial H(p,q)}{\partial p}i\frac{\partial}{\partial q}.
\]
Then both in classical and in quantum statistical mechanics, the state 
is a density operator. The only difference between the classical and 
the quantum case is that in the former case, all operators are diagonal.
In particular, the classical statistical mechanics of macroscopic 
matteris also described by (diagonal) Gibbs states.

As discussed in Part II \cite{Neu.IIfound}, functions of expectations 
satisfy a Hamiltonian dynamics given by a Poisson bracket. It is not 
difficult to show that the Koopman dynamics resulting in this way from
the Koopman Hamiltonian exactly reproduces the classical Hamiltonian 
dynamics of arbitrary systems in which the initial condition is treated 
stochastically. The Koopman dynamics is -- like von Neumann's dynamics 
-- strictly linear in the density matrix. But the resulting dynamics is
highly nonlinear when rewritten as a classical stochastic process.
This is a paradigmatic example for how nonlinearities can naturally 
arise from a purely linear dynamics. 

Because of the Koopman representation, everything said in the following 
about quantum statistical mechanics applies as well to classical 
statistical mechanics.

\subsection{Coarse-graining}\label{ss.coarse}

\nopagebreak
\hfill\parbox[t]{10.8cm}{\footnotesize

{\em Die vorher scheinbar unl\"osbaren Paradoxien der Quantentheorie 
beruh\-ten alle darauf, da{\ss} man diese mit jeder Beobachtung 
notwendig verbundene St\"orung vernachl\"assigt hatte}

\hfill Werner Heisenberg, 1929 \cite[p.495]{Heisenberg1929Naturw}
}

\bigskip

The same system can be studied at different levels of resolution.
When we model a dynamical system classically at high enough resolution,
it must be modeled stochastically since the quantum uncertainties must
be taken into account. But at a lower resolution, one can often neglect
the stochastic part and the system becomes deterministic. If it were 
not so, we could not use any deterministic model at all in physics but 
we often do, with excellent success. 

Coarse-graining explains the gradual emergence of classicality,
due to the law of large numbers to an ever increasing accuracy as
the object size grows. The quantum dynamics changes gradually into 
classical dynamics. The most typical path is through nonequilibrium 
thermodynamics (cf. Subsection \ref{ss.noneq} below). There are also 
intermediate stages modeled by quantum-classical dynamics (see 
Subsection \ref{ss.QCd} below); these are used in situations where the 
quantum regime is important for some degrees of freedom but not for 
others. In fact, there is a wide spectrum of models leading from full 
quantum models over various coarse-grained models to models with a 
fully classical dynamics. One typically selects from this spectrum the 
model that is most tractable computationally given a desired accuracy.

A coarse-grained model is generally determined by singling out a 
vector space $R$ of \bfi{relevant quantities} whose q-expectations are
the variables in the coarse-grained model. If the coarse-grained model
is sensible one can describe a deterministic or stochastic 
\bfi{reduced dynamics} of these variables alone, ignoring all the other
 q-expectations that enter the deterministic Ehrenfest dynamics 
(see Part II \cite[Subsection 2.1]{Neu.IIfound}) 
of the detailed description of the system. These other variables 
therefore become \bfi{hidden variables} that would determine the 
stochastic elements in the reduced stochastic description, or the 
prediction errors in the reduced deterministic description. The hidden 
variables describe the unmodeled \bfi{environment} associated with the 
reduced description.\footnote{
They may be regarded as the hidden variables for which Einstein and 
others searched for so long. Most of them are highly non-local, in 
accordance with Bell's theorem.
The thermal interpretation thus reinstates nonlocal hidden variable 
realism, but -- unlike traditional hidden variable approaches -- 
without introducing additional degrees of freedom into quantum 
mechanics.
} 
Note that the same situation in the reduced description corresponds to 
a multitude of situations of the detailed description, hence each of 
its realizations belongs to different values of the hidden variables 
(the q-expectations in the environment), slightly causing the 
realizations to differ. Thus any coarse-graining results in small 
prediction errors, which usually consist of neglecting experimentally 
inaccessible high frequency effects. These uncontrollable errors are
induced by the variables hidden in the environment and introduce a 
stochastic element in the relation to experiment even when the 
coarse-grained description is deterministic. 

The thermal interpretation claims that this influences the results 
enough to cause all randomness in quantum physics, so that there is no 
need for intrinsic probability as in traditional interpretations of 
quantum mechanics. In particular, it should be sufficient to explain
from the dynamics of the universe the statistical features of 
scattering processes and the temporal instability of unobserved 
superpositions of pure states -- as caused by the neglect of the 
environment. 

\bigskip

To give a concrete example of coarse-graining we mention 
\sca{Jeon \& Yaffe} \cite{JeoY}, who derive the hydrodynamic equations 
from quantum field theory for a real scalar field with cubic and quartic
self-interactions. Implicitly, the thermal interpretation is used, which
allows them to identify field expectations with the classical values of 
the field. 

There are many systems of practical interest where the most slowly 
varying degrees of freedom are treated classically, whereas the most 
rapidly oscillating ones are treated in a quantum way. The resulting
quantum-classical dynamics, discussed in Subsection \ref{ss.QCd} below, 
also constitutes a form of coarse-graining. The approximation of 
fields (with an infinite number of degrees of freedom) by finitely many 
particles is also a form of coarse-graining.

In the context of coarse-graining models given in a Hamiltonian quantum
framework, the \bfi{Dirac--Frenkel variational principle} may be
profitably used for coarse-graining whenever a pure state approximation 
is reasonable. This principle is based on the fact that the integral
\lbeq{e.DF}
I(\psi) = \int \psi(t)^* (i\hbar\partial_t - H) \psi(t) dt
= \int\Big(i\hbar\psi(t)^*\dot\psi(t) - \psi(t)^*H\psi(t)\Big) dt
\eeq
is stationary iff $\psi$ satisfies the time-dependent Schr\"odinger
equation $i\hbar\dot\psi(t) = H\psi(t)$.
Suppose now that a family of pure states $\phi_z$ (depending smoothly 
on a collection $z$ of parameters) is believed to approximate the class 
of states realized in nature we may make the coarse-graining ansatz  
\[
\psi(t)=\phi_{z(t)}
\]
and determine the time-dependent parameters $z(t)$ by finding the 
differential equation for the stationary points of $I(\phi_z)$ varied 
over all smooth functions $z(t)$.
This variational principle was first used by \sca{Dirac} \cite{Dir.var} 
and \sca{Frenkel} \cite{Fre}, and found numerous applications; 
a geometric treatment is given in \sca{Kramer \& Saraceno} \cite{KraS}.

Decoherence (see, e.g., \sca{Schlosshauer} \cite{Schl,Schl.book}) is a
typical phenomenon arising in coarse-grained models of detailed quantum 
systems involving a large environment. It shows that in a suitable 
reduced description, the density operators soon get very close to 
diagonal, recovering after a very short decoherence time a Koopman 
picture of classical mechanics. Thus decoherence provides in principle 
(though only few people think of it in these terms) a reduction of the 
quantum physics of an open system to a highly nonlinear classical 
stochastic process.

For how coarse-graining is done in more general situations given a 
fundamental quantum field theoretic description, see, e.g., 
\sca{Balian} \cite{Bal2}, \sca{Grabert} \cite{Grab}, 
\sca{Rau \& M\"uller} \cite{RauM}. 
In general, once the choice of the resolution of modeling is fixed, 
this fixes the amount of approximation tolerable in the ansatz, and 
hence the necessary list of extensive  quantities. What is necessary 
is not always easy to see but can often be inferred from the practical
success of the resulting coarse-grained model.

\subsection{Chaos, randomness, and quantum measurement}
\label{ss.Balian}

Many coarse-grained models are chaotic. In general, deterministic chaos,
as present in classical mechanics, results in empirical randomness.
For example, the Navier--Stokes equations, used in practice to model 
realistic fluid flow, are well-known to be chaotic. They exhibit 
stochastic features that make up the phenomenon of turbulence.

In the thermal interpretation of quantum physics, empirical randomness 
is also taken to be an emergent feature of deterministic chaos implicit 
in the deterministic dynamics of the Ehrenfest picture discussed in 
Part II \cite{Neu.IIfound}. Since the Ehrenfest dynamics is linear, it 
seems to be strange to consider it chaotic. However, the chaotic 
nature appears once one restricts attention to the macroscopically 
relevant q-expectations, where the influence of the ignored beables
is felt as a stochastic contribution to the effective coarse-grained 
dynamics of the relevant q-expectations. 

To explain the randomness inherent in the measurement of quantum
observables in a qualitative way, the chaoticity of coarse-grained
approximations to equations of motion seems to be sufficient. 
The latter shows how the
deterministic dynamics of the density operator gives rise to stochastic
features at the coarse-grained level. The quantitative derivation of
the stochastic properties is therefore reduced to a problem of quantum
statistical mechanics. 

The dynamics we actually observe is the quantum
dynamics of a more complex system, coarse-grained to a dynamics of
these few degrees of freedom -- at increasing level of coarse-graining
described by Kadanoff--Baym equations, Boltzmann-type kinetic
equations, and hydrodynamic equations such as the Navier--Stokes 
equations. These coarse-grained systems generally behave like classical 
dynamical systems with regimes of highly chaotic motion. 

In general, deterministic chaos manifests itself once one uses a
coarse-grained, locally finite-dimensional parameterization of the 
quantum states.
This leads to an approximation where, except in exactly solvable
systems, the parameters characterizing the state of the universe
(or a selected part of it) change dynamically in a chaotic fashion.

\sca{Zhang \& Feng} \cite{ZhaF} used the Dirac--Frenkel variational
principle introduced in Subsection \ref{ss.coarse}, restricted to group 
coherent states, to get a coarse-grained
system of ordinary differential equations approximating the dynamics of
the q-expectations of macroscopic operators of certain multiparticle
quantum systems. At high resolution, this deterministic dynamics is
highly chaotic. While this study makes quite special assumptions, it
illustrates how although the basic dynamics in quantum physics is 
linear, chaotic motion results once attention is restricted to a 
tractable approximation. This chaoticity is indeed a general feature of 
coarse-graining approximation schemes for the dynamics of 
q-expectations or the associated reduced density functions.
(For a discussion of quantum chaos from a completely different 
perspective see \sca{Peres} \cite[p.353ff]{Per} and the survey by 
\sca{Haake} \cite{Haa2010}.)

\bigskip

According to the thermal interpretation, quantum physics is the basic 
framework for the description of objective reality (including everything
reproducible studied in experimental physics), from the smallest to the 
largest scales. 
In particular, quantum physics must give an account of whatever happens 
in an experiment, when both the equipment and the systems under study 
are modeled on the quantum level.
In experiments probing the foundations of quantum physics, one
customarily observes a small number of field and correlation
degrees of freedom (often simplified in a few particle setting) 
by means of macroscopic equipment.
To model the observation of such a tiny quantum system by a
macroscopic detector one must simply extend the coarse-grained
description of the detector by adding a few additional quantum degrees
of freedom for the measured system, together with the appropriate
interactions. The metastability needed for a reliable quantum detector
(e.g., in a bubble chamber) together with chaoticity then naturally
leads to a random behavior of the individual detection events.

In terms of the thermal interpretation, the \bfi{measurement problem} --
how to show that an experimentally assumed relation between measured 
system and detector results is actually consistent with the quantum 
dynamics --  becomes a precise problem in quantum statistical
mechanics.\footnote{
On the other hand, a somewhat ill-posed, vexing 
measurement problem arises when one insists on the rigid, far too 
idealized framework in which quantum physics was developed historically 
and in which it is typically introduced in textbooks.
} 
Of course, details must be derived in a mathematical manner from the 
theoretical assumptions inherent in the formal core.

A number of recent papers by \sca{Allahverdyan, Balian \& Nieuwenhuizen}
(in the following short AB\&N), reviewed in \sca{Neumaier} 
\cite{Neu.ABNreview}, addressed this issue. 
Here we only discuss AB\&N's paper \cite{AllBN2}, which carefully
analyzed the assumptions regarding the statistical mechanics used that
actually go into the analysis in their long, detailed paper
\cite{AllBN1} of a slightly idealized but on the whole realistic 
measurement process formulated completely in terms of quantum dynamics.

To avoid circularity in their arguments, AB\&N introduce the name 
\bfi{q-expectation value} for $\<A\>:=\Tr \rho A$ considered as a 
formal construct rather than a statistical entity, and similarly (as 
we do in Footnote $^{\ref{f.q}}$\,) q-variance and other q-notions,
to be distinguished from their classical statistical meaning. This
allows them to use the formalism of statistical mechanics without
any reference to prior statistical notions. The statistical
implications are instead derived from the analysis within this formal
framework (together with explicitly specified interpretation rules),
resulting in a derivation of Born's rule and the time scales in
which the implied correlations of microscopic state and measurement
results are dynamically realized, based on a unitary dynamics of the
full quantum system involving the microscopic system, the measurement
device, and a heat bath modeling the environment. 

Most important for the interpretation in \cite{AllBN2} is AB\&N's
''interpretative principle 1'':

\bfi{ABN principle}: If the q-variance of a macroscopic
observable is negligible in relative size its q-expectation value is
identified with the value of the corresponding macroscopic physical
variable, even for an individual system.

This is just a special case of the basic uncertainty principle central
to the thermal interpretation of quantum physics!

\subsection{Gibbs states}\label{ss.Gibbs}

The detailed state of a quantum system can be found with a good
approximation only for fairly stationary sources of very small objects,
of which sufficiently many can be prepared in essentially the same
quantum state. In this case, one can calculate sufficiently many
expectations by averaging over the results of multiple experiments on
these objects, and use these to determine the state via some version of
quantum state tomography \cite{wik.QStomography}. Except in very simple
situations, the result is a mixed state described by a density operator.
Mixed states are necessary also to properly discuss properties of 
subsystems (see Part II \cite{Neu.IIfound}) and for the realistic 
modeling of dissipative quantum systems by equations of Lindblad type 
(\sca{Lindblad} \cite{Lin}). Even for the multi-photon states
used to experimentally check the foundations of quantum physics,
quantum opticians use density operators and not wave functions, since
the latter do not provide the effficiency information required to rule
out loopholes.

Although only a coarse-grained description of a macroscopic system
can be explicitly known, this does not mean that the detailed state
does not exist. {\em The existence of an exact state for large objects 
has always been taken as a metaphysical but unquestioned assumption.}
Even in classical mechanics, it was always impossible to know the exact
state of the solar system with sun, planets, asteroids, and comets 
treated as rigid bodies). But before the advent of quantum mechanics
shattered the classical point of view, its existence was never 
questioned.

Motivated by the above considerations, the thermal interpretation
takes as its ontological basis the density operators, the states
occurring in the statistical mechanics, rather than the pure states
figuring in traditional quantum physics built on top of the concept of 
a wave function.

In the thermal interpretation, all realistic\footnote{
This excludes more idealized states, for example pure states.
All states, including the idealized ones, are obtainable as limits of
Gibbs states. This is because the positive definite density operators
are dense in the set of all density operators, and every positive
definite density operator is a Gibbs state. Indeed, being trace class
and Hermitian, a density operator is self-adjoint, and positive
definiteness implies the existence of the self-adjoint entropy operator
$S=-\kbar \log\rho$, showing that \gzit{e.Gibbs} holds.
In particular, it is experimentally impossible to distinguish between a
pure state and Gibbs states sufficiently close to the pure state.
} 
states are described (as in quantum statistical mechanics) by
\bfi{Gibbs states}, i.e., density operators of the form
\lbeq{e.Gibbs}
\rho:=e^{-S/\kbar},
\eeq
where $\kbar$ the Boltzmann constant and $S$ is a self-adjoint Hermitian
quantity called the \bfi{entropy} of the system in the given state. 
(The traditional entropy is the uncertain value $\<S\>$ of the present 
quantity $S$.) 
Note that a unitary transform $\rho'=U\rho U^*$ of a Gibbs state
by a unitary operator $U$ is again a Gibbs state; indeed, the entropy
of the transformed state is simply $S'=USU^*$. This shows that the 
notion of a Gibbs state is dynamically well-behaved; the von Neumann 
dynamics ensures that we get a consistent evolution of Gibbs states. 

On the level of Gibbs states, the notion of superposition becomes
irrelevant; one cannot superimpose two Gibbs states.
Pure states, where superpositions are relevant, appear only in a limit
where the entropy operator has one dominant eigenvalue and then a
large spectral gap. For example, as we have seen in Part I 
\cite[Subsection 2.2]{Neu.Ifound} of this series of papers, this is 
approximately the case for equilibrium systems where the Hamiltonian 
has a nondegenerate ground state and the temperature is low enough. 
For this one needs a sufficiently tiny system. A system containing a 
screen or a counter is already far too large.

The simplest and perhaps most important case of a Gibbs state is that 
of an \bfi{equilibrium state} of a pure substance, defined by the 
formula
\[
S= (H+PV-\mu N)/T,
\]
where $H$ is the Hamiltonian, $V$ is the system volume, $N$ a 
nonrelativistic number operator, and \bfi{temperature} $T$, 
\bfi{pressure} $P$, and \bfi{chemical potential} $\mu$ are parameters. 
This represents equilibrium states in the form of density operators 
corresponding to grand canonical ensembles,
$\rho=e^{-\beta (H+PV-\mu N)}$, where $\beta=1/\kbar T$.

A derivation of equilibrium thermodynamics in terms of grand canonical 
ensembles in the spirit of the thermal interpretation is given in 
Chapter 10 of \sca{Neumaier \& Westra} \cite{NeuW}.
In this development there is no mention of size. The latter
matters only when one wants to conclude exact thermodynamic results,
since then the thermodynamic limit (infinite volume limit) has to be
taken to reduce the uncertainty to zero.

Temperature $T$, pressure $P$, and chemical potential $\mu$ have no 
simple description in terms of microscopic variables. They figure only 
as a parameter in the expression for the grand canonical phase space 
density $\rho=e^{-(H+PV-\mu N)/\kbar T}$ of the state. But $T$ and $P$ 
are computable from $\rho$ via the thermodynamic formalism of 
statistical mechanics, and hence are beables.

The definition (M) from Subsection \ref{ss.measurement} of what it 
means to measure something therefore applies. More generally, it 
applies (see \sca{Neumaier \& Westra} \cite{NeuW}) 
to arbitrary macroscopic thermal systems in equilibrium, whose 
state is characterized by a collection of finitely many extensive
and intensive thermodynamic variables related by the standard
thermodynamic relations expressed in terms of an equation of state for
the materials making up the thermal system.

In particular, the measurement of temperature and pressure of, say, a 
single brick of iron in equilibrium is a perfectly sensible special 
case of our definition (M) of what it means to measure something.
On the other hand, according to the traditional interpretations, they 
are not even ''observables'' -- although they are observable in any 
meaningful sense of the word!

A realistic system is never exactly in equilibrium, but if it is
sufficiently close to equilibrium, the entropy $S$ is well approximated
by its equilibrium expression $(H+PV-\mu N)/T$. The residual term
$H+PV-\mu N-ST$, which vanishes at equilibrium, contains the detailed
information thrown away in the equilibrium approximation.

\subsection{Nonequilibrium statistical mechanics}\label{ss.noneq}

Unlike in traditional classical or quantum statistical mechanics, the 
density operator \gzit{e.Gibbs} is regarded in the thermal 
interpretation as the complete, exact description of the state, not 
a coarse-grained one. However, one obtaines a coarse-grained reduced
description by replacing the exact $S$ with a suitable approximate $S$ 
given by a more tractable parameterized expression. 

In most coarse-grained models used in statistical mechanics, the form 
assumed for the entropy operator $S$ is a linear combination of 
\bfi{relevant quantities} whose q-expectations define the 
\bfi{extensive variables} of the description. The corresponding 
coefficients are parameters characterizing the particular state of the 
reduced system; they are referred to as the \bfi{intensive variables}
of the description. 
Extensive variables scale linearly with the size of the system
(which might be mass, or volume, or another additive parameter), while 
intensive variables are invariant under a change of system size. 
We do {\em not} use the alternative convention to call extensive any
variable that scales linearly with the system size, and intensive
any variable that is invariant under a change of system size.

If the relevant quantities depend on continuous variables, which is the 
case in nonequilibrium situations, the extensive and intensive
variables become fields depending on the continuum variables used to 
label the subsystems. 
For extensive variables, the integral of their field quantities over 
the label space gives the bulk value of the extensive quantity; thus 
the fields themselves have a natural interpretation as a \bfi{density}. 
For intensive variables, an interpretation as a density is physically 
meaningless; instead, they have a natural interpretation as \bfi{field 
strengths}, sources for \bfi{thermodynamic forces} given by their 
gradients.

In statistical mechanics, we distinguish four nested levels of 
thermal descriptions, depending on whether the system is considered 
to be in global, local, microlocal, or quantum 
equilibrium. The highest and computationally simplest level, 
\bfi{global equilibrium}, is concerned with macroscopic situations 
characterized by finitely many space- and time-independent variables. 

The next level, \bfi{local equilibrium}, treats macroscopic situations 
in a continuum mechanical description, leading, e.g., to the 
Navier--Stokes equations of fluid mechanics. Here the equilibrium 
subsystems are labeled by the space coordinates. Therefore the relevant 
variables are finitely many space- and time-dependent fields. The
entropy operator $S$ becomes time-dependent as is represented as a 
spatial integral
\[
S(t):=\int s(t,x)dx
\]
with a spatial entropy density $s(t,x)$. For a pure monatomic substance,
the latter is in the nonrelativistic case of the form 
\[
s(t,x)= \Big(\eps(t,x)+p(t,x)-\mu(t,x) \rho(t,x)\Big)/T(t,x),
\]
where $\eps(t,x)$ and $\rho(t,x)$ are the internal energy density
and the mass density operators of a quantum field theory whose 
expectations give extensive densities, and $T(t,x)$, $p(t,x)$, and 
$\mu(t,x)$ are intensive coefficient fields defining the local 
temperature, pressure, and chemical potential. (In the relativistic 
case, similar but more involved expressions are used, and the 
identification of temperature and pressure is convention-dependent.)

The next deeper level, \bfi{microlocal}\footnote{
The term microlocal for a phase space dependent analysis is taken
from the literature on partial differential equations; see, e.g., 
\sca{Martinez} \cite{Mar}.
}~ 
\bfi{equilibrium}, treats mesoscopic situations 
in a kinetic description, where the equilibrium subsystems are labeled 
by phase space coordinates. This leads, e.g., to the Boltzmann equation
or the Kadanoff--Baym equations. The relevant variables are now finitely
many fields depending on time, position, and momentum; 
cf. \sca{Balian} \cite{Bal2} and \sca{Rau \& M\"uller} \cite{RauM}.
Now the entropy operator $S$ is represented (in the nonrelativistic 
case) as a phase space integral
\[
S(t):=\int s(t,x,p)dxdp 
\]
with a phase space entropy density $s(t,x,p)$ linearly expressed in
terms of Wigner-trans\-formed operators of a quantum field theory whose 
expectations give extensive phase space densities.

The bottom level is the microscopic regime, where we must consider 
\bfi{quantum equilibrium}. This no longer fits a thermodynamic 
framework but must be described in terms of quantum dynamical 
semigroups and dynamical equations of Lindblad type (\sca{Lindblad}
\cite{Lin}).

Each description level may be considered as a special case of each more
detailed description level. For example, global equilibrium is a 
special case of local equilibrium; the extensive variables in the 
single-phase global equilibrium case have constant densities.

\begin{table}[htb]
\caption{Typical conjugate pairs of thermal variables
and their contribution to the Euler equation. 
The signs are fixed by tradition. (In the gravitational term,
$m$ is the vector with components $m_j$, the mass of a particle of 
kind $j$, $g$ the acceleration of gravity, and $h$ the height.)
}
\label{3.t.}

\begin{center}
{\small
\begin{tabular}{|l|l|l|}
\hline
extensive $X_j$ & intensive $\alpha_j$ 
& contribution $\alpha_jX_j$ \\
\hline
\hline
entropy $S$ & temperature $T$ 
& thermal, $TS$ \\
\hline
particle number $N_j$ & chemical potential $\mu _j$ 
& chemical, $\mu _jN_j$\\
conformation tensor $C$ & relaxation force $R$
& conformational $\sum R_{jk} C^{jk}$\\
\hline
strain $\eps^{jk}$ & stress $\sigma_{jk}$
& elastic, $\sum \sigma_{jk}\eps^{jk}$\\
volume $V$ & pressure $-P$ 
& mechanical, $-PV$ \\
surface $A_S$ & surface tension $\gamma $ 
& mechanical, $\gamma A_S$ \\
length $L$ & tension $J$ 
& mechanical, $JL$ \\
displacement $q$ & force $-F$
& mechanical, $-F\cdot q$\\
momentum $p$ & velocity $v$
& kinetic, $v\cdot p$ \\
angular momentum $J$ & angular velocity $\Omega$
& rotational, $\Omega \cdot J$\\
\hline
charge $Q$ & electric potential $\Phi$
& electrical, $\Phi Q$\\
polarization $P$ & electric field strength $E$ 
& electrical, $E\cdot P$\\
magnetization $M$ & magnetic field strength $B$ 
& magnetical, $B\cdot M$ \\
electromagnetic field $F$ & electromagnetic field strength $-F^s$ 
& electromagnetic, $-\sum F^s_{\mu\nu}F^{\mu\nu}$\\
\hline
mass $M=m\cdot N$ & gravitational potential $gh$
& gravitational, $ghM$ \\
energy-momentum $U$ & metric $g$ 
& gravitational, $\sum g_{\mu\nu}U^{\mu\nu}$ \\
\hline
\end{tabular}
} 
\end{center}
\end{table}

In phenomenological approaches to nonequilibrium thermodynamics, the
entropy operator is written as a linear combination 
\[
S= (H-\sum_j \alpha_j X_j))/T,
\]
of relevant extensive quantities $X_j$, when space is not resolved,
and a corresponding density form when space is resolved to local 
equilibrium. (In microlocal equilibrium, temperature $T$ is no longer 
well-defined, and the lienar combination is written differently.) 
In each case, the relevant quantities are precisely those variables 
that are observed to make a difference in modeling the phenomenon of 
interest.
Table \ref{3.t.} gives typical extensive variables ($S$ and $X_j$), 
their intensive conjugate variables ($T$ and $\alpha_j$), and their 
contribution ($TS$ and $\alpha_j X_j$) to the Euler equation 
\lbeq{e.euler}
H=TS+\sum_j \alpha_j X_j
\eeq
resulting from the definition of the entropy.
Some of the extensive variables and their intensive conjugates are 
vectors or (in elasticity theory, the theory of complex fluids, and in 
the relativistic case) tensors; cf. \sca{Balian} \cite{Bal} for the 
electromagnetic field and \sca{Beris \& Edwards} \cite{BerE}, 
\sca{\"Ottinger} \cite{Oet} for complex fluids.

In general, which quantities need to be considered depends on the 
resolution with which the system is to be modeled -- the higher the 
resolution, the larger the family of extensive quantities. 
Whether we describe bulk matter, surface effects, 
impurities, fatigue, decay, chemical reactions, or transition states, 
-- the general setting remains the same since it is a universal 
approximation scheme, while the number of degrees of freedom 
increases with increasingly detailed models. 

In practice, relevant quantities and corresponding states are assigned 
to real life situations by well-informed judgment concerning the 
behavior of the equipment used. The validity of the assignment is 
experimentally tested by comparing experimental results with the 
chosen mathematical model. The model defines the meaning of the 
concepts: the theory defines what an object is. 

For example, a substance is regarded as an ideal gas if it behaves 
to a satisfactory degree like the mathematical model of an ideal gas
with certain values of temperature, pressure and volume. 
Similarly, a solid is regarded as a crystal if it behaves to a 
satisfactory degree like the mathematical model of a crystal for 
suitable numerical values of the model parameters.

In general, as put by the author of one of the most influential 
textbooks of thermodynamics: {\em ''Operationally, a system is in an 
equilibrium state if its properties are consistently described by 
thermodynamic theory.''} (\label{p.callen}\sca{Callen} \cite[p.15]{Cal})
At first sight, this sounds like a circular definition.
But this is not the case since the formal meaning of ''consistently 
described by thermodynamic theory'' is already known. The operational
definition simply moves it from the domain of theory to the 
domain of reality by defining when a system deserves the designation
''is in an equilibrium state''. In particular, this definition allows
one to determine experimentally whether or not a system is in 
equilibrium.

In general, we know or assume on the basis of past experience, 
claims of manufacturers, etc., that certain materials or machines 
reliably produce states that, to a satisfactory degree for the 
purpose of the experiment or application,
depend only on variables that are accounted for in our theory and 
that are, to a satisfactory degree, either fixed or controllable.
The nominal state of a system can be checked and, if necessary, 
corrected by \bfi{calibration}, using appropriate measurements that
reveal the parameters characterizing the state.

All this is completely independent of a stochastic setting (although
the name 'statistical mechanics' would suggest something different),
and one gets consistent results that compare well with experiment
(if the requirements for the validity of the classical treatment
of certain degrees of freedom are met). Everything can be proved
(at least at the level of typical theoretical physics derivations),
and it has all the beauty and usefulness one might want to have.

\subsection{Conservative mixed quantum-classical dynamics}\label{ss.QCd}

The Koopman representation makes classical systems look quantum.
It is also possible to makes quantum systems look classical. The
resulting quantum-classical dynamics has important applications. 

Since the differences between classical mechanics and quantum mechanics
disappear in the Ehrenfest picture in favor of the common structure of
a classical Hamiltonian dynamics, we can use this framework to mix
classical mechanics and quantum mechanics. The resulting
\bfi{quantum-classical dynamics} is described in many places, e.g., in 
\sca{Peres \& Terno} \cite{PerT.QC},
\sca{Kapral \& Ciccotti} \cite{KapC},
\sca{Prezhdo \& Kisil} \cite{PreK},
\sca{Prezdho} \cite{Pre},
\sca{Breuer \& Petruccione} \cite{BreP.QC}.
The derivation of quantum-classical dynamics from pure quantum
dynamics in these papers follows (under well-understood conditions)
from the principles of statistical mechanics of q-expectations as
embodied in the thermal interpretation, and does not depend on any
measurement issues. Thus it remains valid without any change.

There are many systems of practical interest which are treated in a 
hybrid quantum-classical fashion, where the most slowly varying degrees 
of freedom are treated classically, whereas the most rapidly 
oscillating ones are treated in a quantum way. It is important to have 
an interpretation in which this can be consistently interpreted. 
Any hybrid theory must be interpreted in terms of concepts that have
identical form in classical and in quantum mechanics; otherwise there 
are inevitable conflicts. This is, however, impossible in the
traditional statistical interpretation; there are several theorems
in the literature documenting this \cite{Bou,Sal}.
On the other hand, the thermal interpretation can
cope successfully with this challenge. It is a theory which contains
the classical and the quantum case as two special cases of the
same conceptual framework. In this framework one can therefore discuss
things consistently that lead to puzzles if interpreted either on
a pure classical or on a pure quantum basis, or in some ill-defined
in-between limbo.

The basic equations for a large class of quantum-classical models are,
in the Schr\"odinger picture, the \bfi{Liouville equation}
\lbeq{e.Lio}
    i \hbar \dot\rho = [H(p,q), \rho]
\eeq
and the \bfi{Hamilton equations}
\lbeq{e.Ham}
    \dot q =  \Tr \rho \frac{\partial}{\partial p}H(p,q),~~~
    \dot p=  - \Tr \rho \frac{\partial}{\partial q}H(p,q).
\eeq
Here $q=q(t), p=p(t)$ are classical, time-dependent variables, not
quantum operators, $H(p,q)$ is, for any fixed $p,q$, a linear operator
on some Euclidean space $\Hz$ of smooth wave functions, and 
$\rho=\rho(t)$ is a time-dependent density operator on $\Hz$. 
The sufficiently nice functions of q-expectations
\lbeq{e.trRho}
   \<A(p,q)\> = \Tr \rho A(p,q)
\eeq
where $A$ is a $(p,q)$-dependent operator on a nuclear space, are
classical quantities forming a commutative algebra.
In terms of q-expectations, we have
\[
\dot A = \<H\lp A\>,
\]
and in particular,
\[
\dot q = \Big\<\frac{\partial}{\partial p}H(p,q)\Big\>, ~~~
\dot p= -\Big\<\frac{\partial}{\partial q}H(p,q)\Big\>,
\]
This looks like the original form of the Ehrenfest equations
(Part II \cite[eq. (9)]{Neu.IIfound}), except that on the left hand 
side we have classical variables and no expectations. The expected 
energy $\<H(p,q)\>$ is conserved.

\bigskip

The quantum-classical dynamics preserves the rank of the density $\rho$.
In particular, if $\rho$ has the rank 1 form
\lbeq{esc5}
   \rho = \psi \psi^*
\eeq
at some time, it has at any time the form \gzit{esc5}
with time-dependent $\psi$.
The fact that $\rho$ has trace 1 translates into the statement that
the state vector $\psi$ is normalized to $\psi^*\psi=1$. As discussed 
in detail in Part I \cite[Subsection 2.3]{Neu.Ifound}, the Liouville 
equation \gzit{e.Lio} holds iff the state vector $\psi$, determined by
\gzit{esc5} up to a phase, satisfies -- for a
suitable choice of the phases -- the Schr\"odinger equation
\[
   i \hbar \dot\psi = H(p,q) \psi.
\]
In terms of the state vector, q-expectations now take the familiar form
\[
   \<A(p,q)\> = \psi^* A(p,q) \psi.
\]

In general, q-expectations in the quantum-classical dynamics
are to be interpreted as objects characterizing
a single quantum system, in the sense of the thermal
interpretation, and not as the result of averaging over many
realizations. The quantum-classical dynamics is commonly discussed in
the Schr\"odinger picture, but it is independent of the
picture used. The equivalent Heisenberg dynamics is
\[
    \frac{d}{dt} A =
\frac{\partial A}{\partial q} \Big\<\frac{\partial H}{\partial p}\Big\>
-\frac{\partial A}{\partial p}\Big\<\frac{\partial H}{\partial q}\Big\>
+ \frac{i}{\hbar} [H,A]
\]
where now $\<\cdot\>$ is the fixed Heisenberg state.
From this, one can immediately see that everyhing depends only on
q-expectations by taking expectations in this equation,
\lbeq{e.EhrenfestMixed}
    \frac{d}{dt} \<A\> =
\Big\<\frac{\partial A}{\partial q}\Big\>
\Big\<\frac{\partial H}{\partial p}\Big\>
- \Big\<\frac{\partial A}{\partial p}\Big\>
\Big\<\frac{\partial H}{\partial q}\Big\>
+ \Big\<\frac{i}{\hbar} [H,A]\Big\>.
\eeq
This is now a fully deterministic equation for q-expectations of the
mixed quantum-classical model, considered in the Ehrenfest picture.
This is now the most natural picture, since we still get a Hamiltonian 
description of the form 
\lbeq{e.Edyn}
\frac{d}{dt} \<A\> = \<H\> \lp \<A\>.
\eeq
But now the Lie algebra is the direct product of the Lie algebra of the
classical subsystem and  the Lie algebra of the quantum subsystem. 
This results in a nonlinear dependence on expectations. 
Such nonlinearities are common for reduced descriptions obtained by 
coarse-graining (cf. Subsection \ref{ss.Gibbs} below), both from a pure 
quantum theory or from a classical stochastic theory (in the Koopman 
representation discussed below in Subsection \ref{ss.Koopman}. Since 
quantum-classical systems (at least as they appear in the literature) 
are also coarse-grained descriptions, 
there is nothing surprising in that the same phenomenon occurs.

In the Schr\"odinger picture and the Heisenberg picture, the 
description of a quantum-classical system looks different from that in 
the purely classical and purely quantum case. 

\bigskip

New in quantum-classical systems -- compared to pure quantum dynamics --
is that in the Heisenberg picture, the Heisenberg state occurs 
explicitly in the differential equation for the dynamics. But it does
not take part in the dynamics, as it should be in any good Heisenberg 
picture. The state dependence of the dynamics is not a problem for 
practical applications since the Heisenberg state is fixed anyway by the
experimental setting.

This makes an important difference in the interpretation of the
theory. In contrast to the pure quantum case, there is now a difference
between averaging results of two experiments $\rho_1, \rho_2$ and
the results of a single experiment $\rho$ given by $(\rho_1+\rho_2)/2$.
That, in ordinary quantum theory, the two are indistinguishable
in their statistical properties is a coincidental consequence
of the linearity of the Schr\"odinger equation, and the resulting
state independence of the Heisenberg equation;
it does no longer hold in effective quantum theories
where nonlinearities appear due to a reduced description.
Since quantum-classical systems (at least as they appear
in the literature) also are reduced descriptions,
there is nothing surprising in that the same phenomenon occurs.

Because the dynamics depends on the Heisenberg state, calculating
results by splitting a density at time $t=0$ into a mixture of pure
states no longer makes sense. One gets different evolutions of the
operators in different pure states, and there is no reason why
their combination should at the end give the correct dynamics
of the original density. (And indeed, this will usually fail.)
This splitting is already artificial in pure quantum mechanics
since there is no natural way to tell of which pure states a mixed
state is composed of. But there the splitting happens to be valid
and useful as a calculational tool since the dynamics in the
Heisenberg picture is state independent.

In the quantum-classical case, not even this is possible, so the 
quantum-classical equations have no sensible interpretation in terms 
of mixing pure cases into an ensemble. Thus the quantum-classical 
setting cannot be consistently interpreted in the traditional 
interpretations, where q-expectations have only a statistical meaning. 
But in the thermal interpretation, this is not a problem since 
densities are irreducible objects describing a single quantum system, 
not stochastic entities that make sense only under repetition. Thus in 
the thermal interpretation, the quantum-classical setting is very 
natural.

It is in principle conceivable (though not desirable from the point of
view of simplicity) that the most fundamental description
of nature is truly quantum-classical and not purely quantum.
In the absence of an interpretation with a consistent quantum-classical
setting, this would have been unacceptable, but apart from elegance, 
there are no longer fundamental reasons that would forbid it.

\subsection{Important examples of quantum-classical dynamics}
\label{ss.QCex}

There are many examples of quantum-classical dynamics of practical 
importance.

Probably the most important quantum-classical system is a version of 
the Born--Oppenhei\-mer approximation of molecules, widely used in 
quantum chemistry. Here the nuclei are described in terms of classical 
phase space variables, while the electrons are described quantum 
mechanically by means of a state vector $\psi$ in a Hilbert space of 
antisymmetrized electron wave functions.

A spinning relativistic electron, while having no purely classical
description, can be modeled quantum-classically by classical phase
space variables $p,q$ and a quantum 4-component spin with Hamiltonian
\lbeq{esc6}
    H(p,q) = \alpha \cdot p + \beta m + e V(q)
\eeq
is a $4\times 4$ matrix parameterized by classical 3-vectors $p=p(t)$
and $q=q(t)$,
$\rho =\rho(t)$ is a positive semidefinite $4\times 4$ matrix of
trace 1, and the trace in equation \gzit{e.trRho} is just the trace of a
$4\times 4$ matrix. One gets the equations \gzit{e.Lio} and \gzit{e.Ham}
from Dirac's equation and Ehrenfest's theorem by an approximation 
involving coherent states for position and momentum. This is just a toy 
example; more useful field theoretic quantum-classical versions (see, 
e.g., \sca{G\'erard} et al. \cite{GerMMP}) lead to well-known Vlasov 
equations for $(p,q)$-dependent $4\times 4$ densities, describing a 
fluid of independent classical electrons of the form \gzit{esc6}.

Other examples include the Schr\"odinger-Poisson equations in 
semiconductor modeling 
and the quantum Boltzmann equation with spin represented by $4\times 4$ 
(or in the nonrelativistic case $2\times 2$) matrices parameterized by 
classical phase space variables. (On the other hand, the 
quantum-Boltzmann equation for spin zero is already a purely classical 
equation, since all its dynamical variables are mutually commuting.)

With even more realism, one needs to add to quantum-classical 
descriptions (cf. Subsection \ref{ss.QCd}) a dissipative collision term 
accounting for interactions, and \gzit{esc6} is no longer adequate 
but needs additional stochastic terms.

\section{The relation to traditional interpretations}
\label{s.trad}

\nopagebreak
\hfill\parbox[t]{10.8cm}{\footnotesize

{\em \"Uber die physikalische Interpretation der Formeln sind die 
Meinungen geteilt.}

\hfill Max Born, 1926 \cite[p.803]{Bor1926b}
}

\bigskip

\nopagebreak
\hfill\parbox[t]{10.8cm}{\footnotesize

{\em Das Einzelsystem tr\"agt wirklich die F\"ahigkeit in sich, einem 
bestimmten Me{\ss}vorgang gegen\"uber in verschiedener Weise zu 
reagieren, d.h. verschiedene Me{\ss}werte f\"ur ein und dieselbe 
Zustandsgr\"o{\ss}e zu liefern: welchen, h\"angt vom Zufall ab, oder 
besser wohl von den Phasenbeziehungen zwischen dem System und dem 
Me{\ss}instrument.}

\hfill Erwin Schr\"odinger, 1929 \cite[Vorwort]{Dar}
}

\bigskip

\nopagebreak
\hfill\parbox[t]{10.8cm}{\footnotesize
{\em I reject the basic idea of contemporary statistical quantum 
theory, insofar as I do not believe that this fundamental concept will 
provide a useful basis for the whole of physics.}

\hfill Albert Einstein, 1949 \cite{Einstein1949}
}

\bigskip

\nopagebreak
\hfill\parbox[t]{10.8cm}{\footnotesize

{\em It is usually believed, that the current orthodox theory actually
accounts for the 'nice linear traces' observed in the Wilson chamber 
etc. I think this is a mistake, it does not.}

\hfill Erwin Schr\"odinger, 1958 \cite[p.163]{Schroedinger1958}
}

\bigskip

\nopagebreak
\hfill\parbox[t]{10.8cm}{\footnotesize

{\em Personally I still have this prejudice against indeterminacy in 
basic physics.}

\hfill Paul Dirac, 1972 \cite[p.7]{Dir73}
}

\bigskip

\nopagebreak
\hfill\parbox[t]{10.8cm}{\footnotesize

{\em When it comes to specifying exact details, one discovers that we 
cannot rigorously define what quantum mechanical amplitudes are, what 
it means when it is claimed that 'the universe will collapse with
 such-and-such probability', what and where the observers are, what 
they are made of, and so on. Yet such questions are of extreme 
importance if one wants to check a theory for its self-consistency, by 
studying unitarity, causality, etc.}

\hfill Gerard 't Hooft, 1999 \cite[p.95]{tHoo}
}

\bigskip

\nopagebreak
\hfill\parbox[t]{10.8cm}{\footnotesize

{\em My own conclusion (not universally shared) is that today there is 
no interpretation of quantum mechanics that does not have serious 
flaws.}

\hfill Steven Weinberg, 2013 \cite[p.95]{Wei.QM}
}

\bigskip

From its very beginning in 1926, what turned out to be the formal core 
of quantum mechanics had conflicting interpretations -- initially the
deterministic view of Schr\"odinger and the statistical view of Born.
In 1929, Schr\"odinger conceded the need for a statistical 
interpretation. But the details remained controversial. Today, after 
almost 100 years of interpretation quarrels, the matter is still not 
settled.
As the above quotes show, many of the founders of quantum mechanics 
were never satisfied with the interpretation of quantum mechanics, and
even some of today's Nobel prize winners spend significant effort on 
the interpretation issue.\footnote{
But apparently they did so only after their retirement: While paid they 
researched more important issues and kept -- like most quantum 
physicists -- the foundational issues on the back burner.
} 

A multitude of interpretations of quantum mechanics exist; most of 
them in several variants. We distinguish the following types:

(I) \bfi{Individual interpretations} (such as certain variants of the 
Copenhagen interpretation), where the state of a system is determined by
an individual realization of the system and contains the information 
about everything that can possibly be said about it.

(S) \bfi{Statistical interpretations} (such as the minimal 
interpretation), where the state of a system says (except in special 
cases) nothing about a single system but is only about statistical 
predictions of actual measurements on an ensemble of similarly prepared 
systems.

(K) \bfi{Knowledge interpretations}, where a state says nothing 
objective about the systems modeled, but is only about the subjective 
knowledge of these system.

(O) \bfi{Other interpretations}, where a state consists (as in Bohmian 
mechanics) of more than a state vector or density operator, or is (as 
in many worlds interpretations) by their conception about more than 
actual events recorded in actual experiments.

In the mainstream interpretations of the types (I) and (S), the result 
of a single measurement is -- in contrast to classical mechanics -- 
not even theoretically determined before the measurement is 
done.\footnote{
However, expositions of both views generally prefer to remain vague 
or even silent about this.
} 

As we shall see in this section, the mainstream interpretations may be 
regarded as partial versions of the thermal interpretation. 
In particular, certain puzzling features of both the Copenhagen 
interpretation and the statistical interpretation get their explanation 
through the thermal interpretation of quantum field theory.
We shall see that these peculiar features get their natural 
justification in the realm for which they were created -- the 
statistics of few particle scattering events.

Interpretations of the types (K) and (O) have little in common with the 
thermal interpretation and are not further discussed.

\subsection{The statistical mechanics of definite, discrete events}
\label{ss.discrete}

Generally in physics, invariance and the resulting reproducibility 
determine what counts as an objective property of what:
In 3-dimensional vision, observed length is not a property of an 
observed object by itself but a property of the object and the distance 
from the observer. Extrapolation to zero distance defines an invariant 
objective length.
In relativity, length is no longer a property of an observed object 
and the distance from the observer but a property of the object and 
the distance from and relative speed to the observer. Extrapolation 
to zero distance and zero velocity defines the invariant objective 
length.

Science is about reproducible aspects of our world, and hence not all 
permanent records but only reproducible results count as measurement 
results. This is the main difference between the thermal interpretation 
and traditional interpretations of quantum mechanics.

As a consequence, a measurement of a Hermitian quantity $A$ gives an 
uncertain value approximating the q-expectation $\<A\>$ rather than 
(as tradition wanted to have it) an exact eigenvalue of $A$. 
This difference is most conspicuous in the interpretation of single 
discrete events. Since most single microscopic observations are not 
reproducible they have no scientific value in themselves, and do not 
constitute measurement results.\footnote{
The same holds in classical stochastic models. If die casting is part 
of a stochstic system description, the single die cast tells nothing 
about the state of the model and hence is of no value for the scientific
study of the model. 
} 
Scientific value is, however, in ensembles of such observations, which 
result in approximate measurements of q-probabilities and 
q-expectations.

In the thermal interpretation, the traditional difficulty to show that 
there is always a unique outcome is trivially solved since by 
definition, the outcome of reading a macroscopic quantity is its
expectation value, with negligible uncertainty. Instead we now have a 
new difficulty absent in traditional interpretations: An explanation is 
required why, although fed with a stationary interaction, certain 
detectors record random individual events!

For example, why does a low intensity beam of light produce in a 
photodetector a discrete signal? The uncertain observed value is the 
q-expectation of a photocurrent, which a priori has a continuum of 
possible values. But observed are two clearly different regimes that 
allow one to clearly distinguish between the occurrence and the 
nonoccurence of a detection event. In the thermal interpretation, we do 
not consider the single detector event as a property of the observed 
beam (''a particle arrived through the beam''), since only the 
statistics of an ensemble of detector events (e.g., a Poisson 
distribution of the number of events in some large time interval) is 
reproducible and hence constitutes an objective property of the beam. 
But why these discrete events can be clearly distinguished at all needs
an explanation.

Section 6.6 of the book on open quantum systems by 
\sca{Breuer \& Petruccione} \cite[pp.348--350]{BreP.OQS} (in the 
following short B\&P) addresses this issue. The dynamics of a large 
quantum system, consisting of an observed system and a detector 
observing it, is treated there as a classical dynamical system for the 
density operator with stochastic initial conditions, and reduced by 
appropriate coarse-graining to a classical stochastic equation for the 
coarse-grained stochastic density operator. The derivation is done
using standard assumptions from classical statistical mechanics only, 
in the same way as one would proceed in statistical mechanics for any 
other classical dynamical system.

The detector must include enough of the environment to produce
irreversible results (and hence determines what is read out).
B\&P model the latter by assuming separated time scales and the
validity of the Markov approximation - which hold only if the detector
is big enough to be dissipative. (The latter is typically achieved by
including in the detector a heat bath consisting of an infinite number
of harmonic oscillators.) Since B\&P make these assumptions without
deriving them, their analysis holds for general dissipative detectors.
But -- as always in statistical mechanics -- one must check for any
concrete application that these assumptions are plausible.

In sufficiently idealized settings, these assumptions can actually be
proved rigorously, but this is beyond the scope of the treatment by
B\&P. Rigorous results (without the discussion of selective
measurement but probably sufficient to establish the assumptions used
by B\&P) were first derived by Davies 1974 and later papers with the
same title. See also the detailed survey by \sca{Spohn} \cite{Spo}.

The stochastic equations discussed by B\&P preserves the rank of the 
density operator, and hence can be applied to pure states, where the 
dynamic reduces in general to that of a piecewise deterministic 
stochastic process (PDP), a diffusion process, or a combination of both.
The piecewise deterministic part accounts for the statistics of 
\bfi{discrete events}. 

In the cases treated by B\&P in Chapter 6 (usually for the pure case 
only), the PDP corresponds to photodetection, which measures the 
particle number operator (with a discrete spectrum); the diffusion 
processes correspond to homodyne or heterodyne detection, which measure 
quadratures (with a continuous spectrum).
B\&P obtain the latter from the PDP by a limiting process in the
spirit of the traditional approach treating a continuous spectrum as a
limit of a discrete spectrum.

But although the pointer reading is a position measurement of the
pointer, what is measured about the particle is not its position but
the variable correlated with the pointer reading -- the photon
number or the quadrature. Particle position is as indeterminate as
before. Indeed, investigation of the PDP process shows that the
collapsed states created by the PDP are approximate eigenstates of the
number operator or the quadrature. Thus the PDP can be interpreted in
Copenhagen terms as constituting the repeated measurement of particle
number or quadrature.

For photodetection, one gets at the end a PDP for 
the reduced state vector, only using classical probabilities in the 
whole derivation.  But after everything has been done, the PDP may be 
interpreted in terms of quantum jumps, without having postulated any
irreducible ''collapse'' as in the Copenhagen interpretation 
(cf. Subsection \ref{ss.Copenhagen}). This suggests that, in general, 
that collapse in a single observed system -- in the modern POVM version 
of the von Neumann postulates for quantum dynamics -- is derivable from 
the unitary dynamics of a bigger system under the standard assumptions 
that go into the traditional derivations in classical statistical 
mechanics. 

The arguments show that to go from unitarity to irreversible discrete 
events in Hamiltonian quantum mechanics one does not need to assume 
more than to go from reversibility to irreversibility in Hamiltonian 
classical mechanics -- namely a suitable form of the Markov 
approximation. Statistical assumptions are not needed to make  
pointers acquire a well-defined position or to create photocurrents -- 
the standard dissipation arguments are enough. This gives stochastic
equations for definite macroscopic outcomes.

\subsection{Dissipation, bistability, and Born's rule}\label{ss.Born}

The development of B\&P is mathematical and quantitative but abstract. 
We now provide a qualitative explanation why the discreteness that makes
its appearance in quantum mechanics is actually quite natural, explained
by environment-induced randomness and the associated environment-induced
dissipation. This provides a more intuitive view of how the thermal 
interpretation settles this foundational key issue. (For generalities 
about  environment-induced randomness and dissipation see, e.g., 
the first two chapters of \sca{Calzetta \& Hu} \cite{CalH.book}.)

In general, dissipation in the effective, human time scales dynamics of 
a set of relevant variables is a frequent situation even when the fully 
detailed dynamics is conservative. This effective dissipation is the 
reason underlying the possibility of reduced, coarse-grained 
descriptions whenever there is a separation of time scales for slow
and fast processes. Then one can coarse-grain by eliminating the fast 
modes and obtain a simpler limiting (effectively time-averaged) 
description on the slow manifold, the manifold where all slow motion 
happens (see, e.g., \sca{Lorenz} \cite{Lor}, \sca{Roberts} \cite{Rob}). 
Whenever the slow manifold is disconnected, metastable states of the 
full mainfold decay under uncontrollable (environment-induced) 
perturbations into states in one of the connected components of the 
slow manifold. The components thus label random events selected by 
environmental noise.

For example, consider bending a classical, rotationally symmetric rod 
using a force in the direction of the axis of the rod. If the force 
exceeds the threshold where the straight rod becomes metastable only, 
the rod will bend into a random, but definite direction. The randomness 
arises from the classical Hamiltonian dynamics together with the 
tiniest amount of noise causing a deviation from perfect symmetry.
The same analysis can be made for the dynamics of a metastable inverted
classical pendulum.

Similarly, perturbing in an uncontrollable way a classical bistable 
system arbitrarily little from the intermediate metastable state linking
the two local minima of the potential leads to a tiny random move into
one of the two potential wells. Even the slightest amount of dissipated
energy fixes the selection of the potential well, and more dissipation 
forces the system after a short relaxation time to be very close to 
one of the two minimum position. This is the principle underlying 
the emergence of chemical reactions of molecules (recognizable bound 
states of few atoms) from a multiparticle atomic description in
transition state theory (\sca{H\"anggi} et al. \cite{Haenggi}).

Papers on optical bistability (e.g., \sca{Drummond \& Walls}
\cite{DruW}, \sca{Steyn-Ross \& Gardiner} \cite{SteG}) show how
coarse-grained bistability arises from a quantum model by projecting out
irrelevant degrees of freedom. Any bistable system obtained as a reduced
description from a larger unitary system behaves in the same way.
Thus one expects a few-particle quantum system coupled to a
macroscopic metastable instrument to behave in the same way when (as 
is usual) unstable stationary points are present.

Thus bistability and more general multistability, together with 
dissipation leads, within the accuracy of the approximations involved, 
to the emergence of random discrete events from deterministic dynamics.
The time scale of the emergence of these discrete events is likely to 
be a small multiple of the decoherence time of the system; cf. 
\sca{Schlosshauer} cite{Schl.book}.

\bigskip

The traditional introductory textbook approach to measurement is based
on the concept of ideal measurements -- illustrated with Stern--Gerlach 
experiments, low density double-slit experiments, and the like. These
experiments illustrate an antiquated view of measurement, dating back 
to the time before 1975, when POVMs (see Subsection \ref{ss.event}) 
were still unknown. Until then, quantum measurements used to be 
described solely in terms of ideal statistical measurements. These 
constitute a special case (or for continuum measurements a special 
limiting case) of POVMs where the $P_k$ form a family of 
\bfi{orthogonal projectors}, i.e., linear operators satisfying
\[
P_k^2=P_k=P_k^*,~~~P_jP_k=0\for j\ne k,
\]
to the eigenspaces of a self-adjoint quantity $A$ (or the components of 
a vector $A$ of commuting such quantities) with discrete spectrum 
given by $a_1,a_2,\dots$. We may call a statistical instrument for
measuring $A$ in terms of such a POVM a \bfi{Born instrument}, and the 
instrument is then said to perform an \bfi{ideal measurement} of $A$. 

Ideal measurements of $A$ have quite strong theoretical properties 
since under the stated assumptions, the instrument-based statistical 
average
\[
      \ol {f(A)} = p_1 f(a_1) + p_2 f(a_2) + \dots
\]
agrees for all functions $f$ defined on the spectrum of $A$ with the
model-based value $\<f(A)\>$. 
In an ideal measurement, the relationship between the properties of the
instrument and the properties of the system have a purely correlative
nature, and the rule \gzit{e.Pprob} defining the probabilities reduces 
to the discrete form(e.g., \sca{Drummond \& Walls} \cite{DruW}, 
\sca{Steyn-Ross \& Gardiner} \cite{SteG})  of \bfi{Born's rule}. On the 
other hand, these strong properties are bought at the price of 
idealization, since (unlike more general POVMs) they frequently result 
in effects incompatible with real measurements.

As we saw above, bistability explains the appearance of discrete 
binary events. Once these are given, they provide an ideal binary 
measurement of the statement associated with the event -- giving on a 
single event the result 0 or 1 with a large uncertainty compared with 
the probability it measures. The weak law then implies, as we saw in 
Part II \cite[Subsection 3.5]{Neu.IIfound}, that the relative 
frequencies in sufficiently large samples approximate the probability 
for a positive event, here given by Born's rule for ideal binary 
measurements. 

In particular, for $P_1=\phi\phi^*$, where $\phi$ has norm 1, and 
$P_2 = 1-P$, this covers the case discussed in Part II 
\cite[Subsection 3.4]{Neu.IIfound}, whether a quantum system in the 
pure state $\psi$ responds to a test for state $\phi$, and gives Born's 
squared probability amplitude formula $p = |\phi^*\psi|^2$ for the 
probability of a positive test result. When interpreted as a measure of 
beam intensity, this formula is identical with Malus' law from 1809 
(cf. Subsection \ref{ss.qubit}).

\subsection{The Copenhagen interpretation}\label{ss.Copenhagen}

\nopagebreak
\hfill\parbox[t]{10.8cm}{\footnotesize

{\em The concept of observation is in so far arbitrary as it depends 
upon which objects are included in the system to be observed. Ultimately
every observation can of course be reduced to our sense perceptions. 
The circumstance, however, that in interpreting observations use has 
always to be made of theoretical notions, entails that for every 
particular case it is a question of convenience at what point the 
concept of observation involving the quantum postulate with its inherent
'irrationality' is brought in.}

\hfill Niels Bohr, 1927 \cite[p.580]{Boh1927}
}

\bigskip

\nopagebreak
\hfill\parbox[t]{10.8cm}{\footnotesize

{\em Um zur Beobachtung zu gelangen, muss man also irgendwo ein 
Teilsystem aus der Welt ausschneiden und \"uber dieses Teilsystem eben 
'Aussagen' oder 'Beobachtungen' machen. Dadurch zerst\"ort man dort den 
feinen Zusammenhang der Erscheinungen und an der Stelle, wo wir den 
Schnitt zwischen dem zu beobachtenden System einerseits, dem Beobachter 
und seinen Apparaten andererseits machen, m\"ussen wir Schwierigkeiten 
f\"ur unsere Ansschauung erwarten. [...] Jede Beobachtung teilt in 
gewisser Weise die Welt ein in bekannte und unbekannte oder besser: 
mehr oder weniger genau bekannte Gr\"ossen.}

\hfill Werner Heisenberg, 1927 \cite[p.593f]{Hei.Como}
}

\bigskip

\nopagebreak
\hfill\parbox[t]{10.8cm}{\footnotesize

{\em wir m\"ussen die Welt immer in zwei Teile teilen, der eine ist das 
beobachtete System, der andere der Beobachter. In der ersteren k\"onnen 
wir alle physikalischen Prozesse (prinzipiell wenigstens) beliebig 
genau verfolgen, in der letzteren ist dies sinnlos. Die Grenze zwischen 
beiden ist weitgehend willk\"urlich}

\hfill John von Neumann, 1932 \cite[p.223f]{vNeu.book}
}

\nopagebreak
\hfill\parbox[t]{10.8cm}{\footnotesize

{\em Aus diesem Zwiespalt ergibt sich die Notwendigkeit, bei der 
Beschreibung atomarer Vorg\"ange einen Schnitt zu ziehen zwischen den 
Me{\ss}apparaten des Beobachters, die mit den klassischen Begriffen 
beschrieben werden, und dem Beobachtungsobjekt, dessen Verhalten 
durch eine Wellenfunktion dargestellt wird. W\"ahrend nun sowohl auf 
der einen Seite des Schnittes, die zum Beobachter f\"uhrt, 
wie auf der anderen, die den Gegenstand der Beobachtung enth\"alt, 
alle Zusammenh\"ange scharf determiniert sind -- hier durch die 
Gesetze der klassischen Physik, dort durch die Differentialgleichungen 
der Quantenmechanik --, \"au{\ss}ert sich die Existenz des Schnittes
doch im Auftreten statistischer Zusammenh\"ange. An der Stelle des 
Schnittes mu{\ss} n\"amlich die Wirkung des Beobachtungsmittels auf den 
zu beobachtenden Gegenstand als eine teilweise unkontroIlierbare 
St\"o\-rung aufgefa{\ss}t werden. [...]
Entscheidend ist hierbei insbesondere, da{\ss} die Lage des Schnittes --
d.h. die Frage, welche Gegenst\"ande mit zum Beobachtungsmittel
und welche mit zum Beobachtungs\-objekt gerechnet werden -- f\"ur die 
Formulierung der Naturgesetze gleichg\"ultig ist.}

\hfill Werner Heisenberg, 1934 \cite[p.670f]{Heisenberg1934Naturw}
}

\bigskip

The \bfi{Copenhagen interpretation} is the interpretation of quantum
mechanics first expressed in 1927 by Bohr and Heisenberg. Until 1970, 
it has been (in various variants) the almost generally accepted 
interpretation though there is no document defining it precisely; 
its contents was stated in varying ways depending on the occasion.
One of the probable reasons is that it had sufficient definiteness 
to guide theory, experiment, and their relationship, and was at the 
same time sufficiently vague that it allowed each user to make sense of 
its paradoxical features in a personal, subjective way. 

In our classification of interpretations of quantum mechanics, the 
Copenhagen interpretation belongs to type (I); the term 'knowledge' 
used first by \sca{Heisenberg} \cite{Hei1927} was not understood in the 
subjective way used in (K) but as the objective (through thought 
experiments theoretically accessible) knowledge of what is real and in 
principle observable about the system, whether observed or not.

\bigskip

One important feature of the Copenhagen interpretation is the so-called
\bfi{Heisenberg cut}, first described by \sca{Heisenberg} 
\cite{Hei.Como} and \sca{Bohr} \cite[pp.580,584]{Boh1927} -- the 
artificial splitting of the world into a quantum domain and a classical 
domain. \sca{von Neumann} \cite[p.223f]{vNeu.book} showed that this cut 
can be places fairly freely without affecting the main conclusions.

While adequate for microscopic systems, the concept of a necessary cut 
fails systematically for sufficiently large systems. For example, as 
all measurements are done within the solar system, it excludes treating 
the solar system as a quantum system (e.g, measuring the mass of the 
earth).

As mentioned already in Subsection \ref{ss.Gibbs}, the thermal 
interpretation nowhere imposes a cut between microscopic an macroscopic.
It is not needed: 
A paper by \sca{Jeon \& Yaffe} \cite{JeoY} derives the hydrodynamic
equations from quantum field theory without assuming a Heisenberg cut.
Only the thermal interpretation is (implicitly) invoked, which allows 
them to identify field expectations with the classical values of fields.

According to the thermal interpretation, classical physics appears 
gradually as systems become more macroscopic.
In continuation of the discussion in Subsection \ref{ss.macro} we call
a quantum system whose relevant quantities have a negligible 
uncertainty a \bfi{classical system}. It is typically described by 
nonequilibrium thermodynamics, as deduced from quantum statistical
mechanics; see Subsection \ref{ss.noneq} below. Thus a classical system 
is still quantum mechanical when modeled in full detail, but only the 
macroscopic variables modeled by statistical mechanics are deemed to 
be relevant. Thus the thermal interpretation leads to a gradual change 
from quantum to classical as the system gets larger and the uncertainty 
of the collection of relevant quantities decreases.

But the thermal interpretation realizes a modified version of the 
Heisenberg cut as the choice of relevent variables in the coarse-grained
description, which defines the split between system and environment.
According to the thermal interpretation, there is no sharp cut but a
 smooth fuzzy boundary, of the same kind as the boundary between the 
Earth's atmosphere and interplanetary space. The bigger one makes the 
instrument the more classical it becomes and the more accurate become 
the pointer positions. In place of deciding where to place the 
Heisenberg cut we now have to decide at which level of description the 
$O(N^{-1/2})$ corrections can be neglected. This is a decision just like
the decision of whether or not to include into the classical description
of a pendulum the surrounding air and the way it is suspended, or
whether taking it into account with a damping term, or even neglecting 
that as well, is enough.

In quantum-classical approximations of a qauntum system, the Heisenberg 
cut is explicitly modeled by allowing for both classica and quantum 
degrees of freedom, neglecting only irrelevant variables. As we have 
seen, the quantum-classical description naturally fits the thermal 
interpretation since q-expectations occur explicitly in the dynamics.

\bigskip

\nopagebreak
\hfill\parbox[t]{10.8cm}{\footnotesize

{\em Die 'Bahn' entsteht erst dadurch, da{\ss} wir sie beobachten}

\hfill Werner Heisenberg, 1927 \cite[p.185]{Hei1927}
}

\bigskip

Another puzzling feature of the Copenhagen interpretation is that an 
individual few particle system has \bfi{no definite properties} before 
it is observed. Taking the Copenhagen interpretation as an irreducible 
description of the nature of things leaves one puzzled how then the 
observing instrument can be informed about what it observes -- being 
virtually nonexistent before the act of observation: In his 1927 paper 
famous for the uncertainty relation, Heisenberg asserted that the path 
(of a particle) is created only through the act of observation. Thus
the observation creates the properties. But it must be created by 
something to be observed! The thermal interpretation gives the
natural answer that this happens because the fields provide the
information about what is there to cause the detectors to respond,
so that something is observed: When measured, particles appear as 
detection events created by the detector and mediated by fields 
(cf. property (P) from Part I \cite[Subsection 4.4]{Neu.Ifound}.

In photodetection, tradition takes the individual detection results too 
seriously and dogmatically\footnote{
There is no way to test this assumption empirically.
} 
interprets the random counting events as signals of single photons 
arriving, with all the spooky problems associated with this view. 
In contrast, the thermal interpretation treats it (as Stokes would have 
done it in 1852) as a very uncertain measurement of energy density. 
Then the particle aspect completely disappears. This is an advantage 
since, as we have seen in Subsection \ref{ss.geomOpt}, it is difficult 
to specify -- even informally -- a particle picture at finite times in 
terms of the underlying relativistic quantum field description. How to 
do this with some degree of mathematical precision is an unsolved 
problem.

\bigskip

\nopagebreak
\hfill\parbox[t]{10.8cm}{\footnotesize

{\em Fragen wir also nicht, wo ist ein Teilchen genau, sondern 
begn\"ugen wir uns, zu wissen, da{\ss} es in einem bestimmten 
gr\"o{\ss}eren Raumteil ist: dann verschwindet der Widerspruch zwischen 
Wellen und Corpus\-culartheorie.}

\hfill Max Born, 1929 \cite[p.116]{Born1929Naturw}
}

\bigskip

There is no doubt that while an atom is in an ion trap it has a 
definite but uncertain position. We know it is there and can check 
efficiently the duration of its presence. Indeed, in order to be able 
to do experiments with single atoms at all we need to know that they are
there! In the Copenhagen interpretation, this knowledge was outside the 
quantum domain, on the classical side of the Heisenberg cut. 
The thermal interpretation preserves the reality of atoms being 
somewhere reasonably well localized while rejecting the idealization
(made in classical point mechanics) that this position is given by an 
exact 3-dimensional real vector. This assumption leads to pardoxes both 
in classical electrodynamics and in certain quantum mechanics 
experiments. In the thermal interpretation it is avoided from the start,
since all quantities come with their intrinsic uncertainty. 

A ''particle trace'' on a photograph is also measurable. Tradition 
postulates that a corresponding particle existed that left this trace.
But this statement is not experimentally refutable by any means, hence 
is a metaphysical assumption. Assuming it we can \bfi{infer} the 
uncertain position and momentum of a  particle that was supposed to 
create it at the time of its creation. But we might also argue as in
Part I \cite[Section 4.4]{Neu.Ifound} and declare in analogy with the 
bullet experiment discussed there the tracks as a meassure of impact 
quality, not associated with any particle!

\sca{Grassi} \cite{Gra} and \sca{Jeon \& Heinz} \cite[Section 5.3]{JeoH}
(and many others) treat (in line with the thermal picture) interacting 
elementary ''particles'' not as particles but as quantum fluids; only 
their freeze-out in scattering experiments produces particle-like 
objects, in a way more or less analogous to how condensed droplets 
appear in saturated liquids. However, the quantization introduces a 
discrete element into the quantum numbers (and hence the number and 
distribution) of the resulting droplets, measurable as impact events or 
particle tracks.

\bigskip

\nopagebreak
\hfill\parbox[t]{10.8cm}{\footnotesize

{\em Jede Ortsbestimmung reduziert also das Wellenpaket wieder auf 
seine urspr\"ungliche Gr\"o{\ss}e
}

\hfill Werner Heisenberg, 1927 \cite[p.186]{Hei1927}
}

\bigskip

\nopagebreak
\hfill\parbox[t]{10.8cm}{\footnotesize

{\em The state of the system after the observation must be an
eigenstate of [the observable] $\alpha$, since the result of a
measurement of $\alpha$ for this state must be a certainty.
}

\hfill Paul Dirac, 1930 \cite[p.49]{Dir}
}

\bigskip

A third significant feature of the Copenhagen interpretation is the 
so-called \bfi{collapse} of the wave function upon measurement, 
introduced in 1927 by \sca{Heisenberg} \cite{Hei1927}. 
Collapse to what is controversial. The authoritative 2007 book by 
\sca{Schlosshauer} \cite{Schl.book}, takes the ''jump into an 
eigenstate'' postulated by Dirac in the above quote to be part of what 
he calls the ''standard interpretation'' of quantum mechanics. On the 
other hand, the older but also authoritative 1977 book by 
\sca{Landau \& Lifschitz} \cite{LL.3} explicitly remarks in the 
discussion in Section 7 that the state after the measurement is in 
general not an eigenstate. 

The collapse requirement contradicts the unitary evolution of pure 
states through the Schr\"odinger equation (which is the mode of change 
of a closed system, hence in the absence of a measurement), and depends 
on a not further detailed notion of measurement residing on the 
classical side of the cut. What happens to the state while the 
experiment is in progress but not complete is not specified. This makes 
the Copenhagen interpretation an incomplete description of the full 
temporal behavior of the state.

This incompleteness is a sign that we actually deal with a 
coarse-grained, reduced description. In such a reduced description,
the description of the state of a particle is different before and 
after it passes a filter (polarizer, magnet, double slit, etc.). 
The new information that the particle passed the filter requires
a different description analogous to that responsible for the use of 
classical conditional probability when additional information arrives.
In the quantum case, this is modeled by the collapse of the wave 
function. Landau's general case is the one modeled by event-based 
filters (Subsection \ref{ss.event}), while Dirac's situation is modeled 
by the special case where the filter operators $R_k$ are the spectral 
projectors of an ideal measurement (Subsection \ref{ss.Born}).

In the thermal interpretation, collapse results from coarse-graining
when the latter produces a reduced stochastic description in the form 
of a PDP (see Subsection \ref{ss.discrete}). For example, we had seen 
in Subsection \ref{ss.QCd} that quantum-classical models may result 
from coarse-graining, and \sca{Bonilla \& Guinea} \cite{BonG} give an 
explicit quantum-classical model that exhibits chaos and collapse.

Understanding that collapse comes from coarse-graining is a similar 
insight as that friction comes from coarse graining, an insight familiar
from classical mechanics treated in the Markov approximation for a few 
relevant quantities.
In both cases, the insight bridges the difference in the dynamics of an 
isolated system and that on an open system obtained by hiding the 
environment, turning it into a source of stochatic events. 
The explanation by coarse graining is in both cases fully quantitative 
and consistent with experiment, hence has all the features a good 
scientific explanation should have.

\subsection{The minimal interpretation}

\nopagebreak
\hfill\parbox[t]{10.8cm}{\footnotesize

{\em I reject the basic idea of contemporary statistical quantum theory,
insofar as I do not believe that this fundamental concept will provide 
a useful basis for the whole of physics.
[...]\\
One arrives at very implausible theoretical conceptions, if one 
attempts to maintain the thesis that the statistical quantum theory is 
in principle capable of producing a complete description of an 
individual physical system. On the other hand, those difficulties of 
theoretical interpretation disappear, if one views the 
quantum-mechanical description as the description of ensembles of 
systems.
[...]\\
Within the framework of statistical quantum theory there is no such 
thing as a complete description of the individual system.
[...]\\
it appears unavoidable to look elsewhere for a complete description of 
the individual system.
[...]\\
I am rather firmly convinced that the development of theoretical physics
will be of this type; but the path will be lengthy and difficult.
[...]\\
the expectation that the adequate formulation of the universal laws 
involves the use of all conceptual elements which are necessary for a 
complete description, is more natural.
[...]\\
If it should be possible to move forward to a complete description, it 
is likely that the laws would represent relations among all the 
conceptual elements of this description which, per se, have nothing to 
do with statistics.
}

\hfill Albert Einstein, 1949 \cite{Einstein1949}
}

\bigskip

\nopagebreak
\hfill\parbox[t]{10.8cm}{\footnotesize

{\em The Statistical Interpretation, according to which a pure state 
(and hence also a general state) provides a description of certain 
statistical properties of an ensemble of similarily prepared systems.
[...]\\
In general, quantum theory predicts nothing which is relevant to a 
single measurement (excluding strict conservation laws like those of 
charge, energy, or momentum), and the result of a calculation pertains 
directly to an ensemble of similar measurements. For example, a single 
scattering experiment consists in shooting a single particle at a 
target and measuring its angle of scatter. Quantum theory does not deal 
with such an experiment, but rather with the statistical distribution 
(the differential cross section) of the results of an ensemble of 
similar experiments. 
}

\hfill Leslie Ballentine, 1970 \cite[p.360f]{Ball1970}
}

\bigskip

\bigskip

\nopagebreak
\hfill\parbox[t]{10.8cm}{\footnotesize

{\em We can now define the scope of quantum theory: In a strict sense, 
quantum theory is a set of rules allowing the computation of 
probabilities for the outcomes of tests which follow specified 
preparations.
[...]\\
The above strict definition of quantum theory (a set of rules for 
computing the probabilities of macroscopic events) is not the way it is 
understood by most practicing physicists. They would rather say that 
quantum theory is used to compute the properties of microscopic objects,
for example the energy-levels and cross-sections of atoms and nuclei. 
The theory can also explain some properties of bulk matter, such as the 
specific heat of solids or the electric conductivity of metals -- 
whenever these macroscopic properties can be derived from those of the
microscopic constituents. Despite this uncontested success, the 
epistemological meaning of quantum theory is fraught with controversy, 
perhaps because it is formulated in a language where familiar words are 
given unfamiliar meanings. Do these microscopic objects -- electrons, 
photons, etc. -- really exist, or are they only a convenient fiction 
introduced to help our reasoning, by supplying intuitive models in 
circumstances where ordinary intuition is useless?
}

\hfill Asher Peres, 2002 \cite[p.13]{Peres}
}

\bigskip

That quantum mechanical states should be interpreted statistically goes 
back to 1926. \sca{Born} \cite{Bor1926a,Bor1926b} had shown how the 
known statistical properties of scattering events is in some sense 
consistent with the deterministic Schr\"odinger equation and can be 
derived from it assuming a statistical interpretation of the wave 
function. This earned him later a Nobel prize. But in Born's view, 
particles existed (as beables) always in joint eigenstates of 
Hamiltonian and momentum that were modified discontinuously by
random \bfi{quantum jumps}. In this way, the exact validity of the 
conservation laws of energy and momentum could be asserted.

Statistical interpretations in the precise sense (S) defined 
at the top of Section \ref{s.trad} have their beginnings with 
\sca{Weyl} \cite{Weyl1927} and were discussed extensively by 
\sca{Ballentine} \cite{Ball1970}, who contrasted it to the Copenhagen 
interpretation. In these statistical interpretations, a single (few or 
many particle) system has no state. Instead, the state is a property of 
the ensemble; one only talks about the prepared and observed properties 
of a population of experiments making up the ensemble. 
Equivalently, the preparation procedure (which defines the ensemble on 
which measurements are performed) has, or defines, a state.

The \bfi{minimal interpretation} is a statistical interpretation (S)
augmented by the additional stipulation that quantum mechanics is
completely silent about a single system, and hence says nothing at all 
about it.\footnote{
True minimality is rare. \sca{Einstein} \cite{ Einstein1949} finds only 
the minimal interpretation consistent, but takes this as a limitation
of quantum mechanics and expects the existence of a deeper underlying 
deterministic description. \sca{Ballentine} \cite{Ball1970} is not 
minimal throughout (despite an attempt to be so), as he assumes (p.361) 
that definite positions exist: {\it ''the Statistical Interpretation 
considers a particle to always be at some position in space, each 
position being realized with relative frequency $|\psi(r)|^2$ in an 
ensemble of similarily prepared experiments.''}
Even the most consequent book by \sca{Peres} \sca{Peres} makes an 
exception at the very end (p.424f):
{\it ''This would cause no conceptual difficulty with quantum theory if 
the Moon, the planets, the interstellar atoms, etc., had a well defined 
state $\rho$. However, I have insisted throughout this book that $\rho$ 
is not a property of an individual system, but represents the procedure 
for preparing an ensemble of such systems. How shall we describe 
situations that have no preparer? [...] Thus, a macroscopic object 
effectively [...] mimics, with a good approximation, a statistical 
ensemble. [...] You must have noted the difference between the present 
pragmatic approach and the dogmas held in the early chapters of this 
book.''}
} 


According to \sca{Ballentine} \cite[p.366]{Ball1970}, for a consistent 
statistical interpretation, the notion of \bfi{preparation} must be
clearly  distinguished from that of \bfi{measurement}:
{\it ''State preparation refers to any procedure which will yield a 
statistically reproducible ensemble of systems. The concept of state in 
quantum theory can be considered operationally as an abbreviation for a
description of the state preparation procedure. Of course there may be 
more than one experimental procedure which yields the same statistical 
ensemble, i.e., the same state. An important special case (which is
sometimes incorrectly identified with measurement) is a filtering 
operation, which ensures that if a system passes through the filter it 
must immediately afterward have a value of some particular observable 
within a restricted range of its eigenvalue spectrum. On the other hand,
measurement of some quantity $E$ for an individual system means an 
interaction between the system and a suitable apparatus, so that we may
infer the value of $R$ (within some finite limits of accuracy) which
the system had immediately before the interaction (or the value of $R$ 
which the system would have had if it had not interacted, allowing for 
the possibility that the interaction will disturb the system).''}
The \bfi{filtering} mentioned replaces the collapse in the Copenhagen 
interpretation, and serves the same purpose. In the thermal 
interpretation, it is modeled by event-based filters (Subsection 
\ref{ss.event}).

The thermal interpretation of the situation described is that the 
preparation defines a state of the quantum fields present in the 
description. Their interaction with the detector produces observable 
events, whose statistics measures properties of the fields. 
In principle, quantum tomography (see Subsection \ref{ss.event}) can 
be used to calibrate sufficiently stationary unknown sources so that 
one can be sure which state they prepare in which setting.
Knowing what was prepared and how to control it systematically, 
one can collect event statistics for new experimental settings and 
establish on the basis of the resulting experimental evidence a 
relation \gzit{e.rhoDS} between measurement results and properties of 
the system measured.

The single systems that allegedly travel from the source to the 
detector (but according to the minimal interpretation without any 
quantum properties) never enter the description, hence one cannot (and 
need not) say anything about these. 

\bigskip

In a statistical interpretation, all statements claimed about single 
quantum systems are non-minimal. In particular, unlike the thermal 
interpretation, the minimal interpretation does not address the 
foundational problems posed by the ensembles of equilibrium 
thermodynamics (cf. Part II \cite[Subsection 2.4]{Neu.IIfound}). 
Indeed, \sca{Ballentine} \cite[p.361]{Ball1970} writes:
{\it ''The ensembles contemplated here are different in
principle from those used in statistical thermodynamics,
where we employ a, representative ensemble for calculations,
but the result of a calculation may be compared
with a measurement on a single system. [...]  
Because the ensemble in the statistical interpretation
''is not merely a representative or calculational device, but rather it 
can and must be realized experimentally, it does not inject into 
quantum theory the same conceptual problems posed in statistical 
thermodynamics.''}

\section{Conclusion}

\nopagebreak
\hfill\parbox[t]{10.8cm}{\footnotesize

{\em Nous tenons la m\'ecanique des quanta pour une th\'eorie
compl\`ete, dont les hypoth\`eses fondamentales physiques et
math\'ematiques ne sont plus susceptibles de modification. 
}

\hfill Max Born, Werner Heisenberg, 1927
                                        \cite[p.178]{BornHeisenberg1928}
}

\bigskip

\nopagebreak
\hfill\parbox[t]{10.8cm}{\footnotesize

{\em According to the present quantum mechanics, the probability
interpretation, the interpretation which was championed by Bohr,
is the correct one. But still, Einstein did have a point.
He believed that, as he put it, the good God does not play with dice.
He believed that basically physics should be of a deterministic
character.\\
And, I think it might turn out that ultimately Einstein will
be proved right, because the present form of quantum mechanics
should not be considered as the final form. [...]
And I think that it is quite likely that at some future time we
may get an improved quantum mechanics in which there will be a
return to determinism and which will, therefore, justify the
Einstein point of view.}

\hfill Paul Dirac, 1975 \cite{Dir}
}

\bigskip

According to the thermal interpretation, quantum physics is the basic 
framework for the description of objective reality (including everything
reproducible studied in experimental physics), from the smallest to the 
largest scales. Classical descriptions are regarded as limiting cases 
when Planck's constant $\hbar$ can be set to zero without significant 
loss of quality of the resulting models. The measurement of a Hermitian 
quantity $A$ is regarded as giving an uncertain value approximating the 
q-expectation $\<A\>$ rather than (as tradition wanted to have it) as 
an exact revelation of an eigenvalue of $A$. Single observations of 
microscopic systems are (except under special circumstances) very 
uncertain measurements only.

It seems that without having to introduce any change in the formal 
apparatus of quantum physics, the deterministic dynamics of the complete
collection of q-expectations constructible from quantum fields, when
restricted to the set of macroscopically relevant ones, already gives 
rise to all the stochastic features observed in practice.

The thermal interpretation  

\pt
is inspired by what physicists actually do rather than what they say. 
It is the interpretation that people actually work with in the 
applications (as contrasted to work on the foundations themselves), 
even when they pay lipservice to another interpretation.

\pt
treats detection events as a statistical measurement of particle beam 
intensity.

\pt
claims that the particle concept is only asymptotically valid, under 
conditions where particles are essentially free.

\pt 
claims that the unmodeled environment influences the results enough to 
cause all randomness in quantum physics.

\pt
allows one to derive Born's rule for scattering and in the limit of 
ideal measurements; but in general, only part of Born's rule holds 
exactly: Whenever a quantity $A$ with zero uncertainty is measured 
exactly, its value is an eigenvalue of $A$.

\pt
has no explicit collapse -- the latter emerges approximately in
non-isolated subsystems.

\pt
has no split between classical and quantum mechanics -- the former
emerges naturally as the macroscopic limit of the latter.

\pt
has no split between classical and quantum mechanics -- the former
emerges naturally as the macroscopic limit of the latter;

\pt
explains the peculiar features of the Copenhagen interpretation 
(lacking realism between measurements) and the minimal statistical 
interpretation (lacking realism for the single case) in the microscopic 
domain where these interpretations apply.

\pt 
paints a deterministic picture of quantum physics in which God does not 
play dice. It only seems so to us mortals because of our limited 
resolution capacity and since we have access to a limited part of the 
universe only.

\bigskip

In terms of the thermal interpretation, the measurement problem turns 
from a philosophical riddle into a scientific problem in the domain of 
quantum statistical mechanics, namely how the quantum dynamics 
correlates macroscopic readings from an instrument with properties of 
the state of a measured microscopic system. 

While the present paper shows that in principle this is enough to 
resolve the riddles of quantum mechanics, a number of detailed questions
remain open:

\pt 
The measurement principle (MP) from Subsection \ref{ss.measurement}
demands that any instrument for measuring a quantity $A$ has an 
uncertainty $\Delta a\ge \sigma_A$. It is an open problem how to prove 
this from the statistical mechanics of measurement models.

\pt 
The derivation of a piecewise deterministic stochastic process (PDP)
by \sca{Breuer \& Petruccione} \cite{BreP.OQS} suggests that, in 
general, that collapse in a single observed system -- in the modern 
POVM version of the corresponding von Neumann postulate for quantum 
dynamics -- is derivable from the unitary dynamics of a bigger system 
under the standard assumptions that go into the traditional derivations 
in classical statistical mechanics. It would be desirable to have
a direct argument for this not dependent on a statistical approach.

\pt 
It should be possible to show in quantitative detail how position 
loses its parameter status and becomes uncertain when going from the 
relativistic quantum field description of a beam to a corresponding 
quantum mechanical description of a sequence of particles moving along 
the beam.

\pt 
It should be sufficient to explain from the dynamics of the 
universe the statistical features of scattering processes and the 
temporal instability of unobserved superpositions of pure states -- 
as caused by the neglect of the environment. 

The experimental evidence for the truth of all this is already there.
Thus, unlike traditional interpretations, the thermal interpretation is 
an interpretation of quantum physics that is in principle refutable by 
theoretical arguments.

\bigskip

\end{document}